\documentclass[12pt,preprint]{aastex}
\begin{document}
\def\ujy     {{$\mu Jy$}}
\def\Lya     {{Ly$\alpha$ }}

\title{Rest-frame Ultraviolet-to-Optical Properties of Galaxies at $z\approx 6$ and 5
in the Hubble Ultra Deep Field: from Hubble to Spitzer}

\author{Haojing Yan\altaffilmark{1},
Mark Dickinson\altaffilmark{2},
Daniel Stern\altaffilmark{3},
Peter R. M. Eisenhardt\altaffilmark{3},
Ranga-Ram Chary\altaffilmark{1},
Mauro Giavalisco\altaffilmark{4},
Henry C. Ferguson\altaffilmark{4},
Stefano Casertano\altaffilmark{4},
Christopher J. Conselice\altaffilmark{5},
Casey Papovich\altaffilmark{6},
William T. Reach\altaffilmark{1},
Norman Grogin\altaffilmark{4},
Leonidas A. Moustakas\altaffilmark{3},
Masami Ouchi\altaffilmark{4}
}

\altaffiltext{1} {Spitzer Science Center, California Institute of Technology,
MS 220-6, Pasadena, CA 91125; yhj@ipac.caltech.edu}
\altaffiltext{2} {National Optical Astronomy Observatory, 950 N. Cherry St.,
Tucson, AZ 85719}
\altaffiltext{3} {Jet Propulsion Laboratory, 4800 Oak Grove Dr., 
Pasadena, CA 91109}
\altaffiltext{4} {Space Telescope Science Institute, 3700 San Martin Dr.,
Baltimore, MD 21218}
\altaffiltext{5} {Department of Astronomy, California Institute of Technology, Mail Code 105-24, Pasadena, CA 91125}
\altaffiltext{6} {Steward Observatory, University of Arizona, 933 North Cherry Av., Tucson, AZ 85721}

\begin{abstract}

   We use data from the first epoch of observations with the Infrared Array
Camera (IRAC) on the {\it Spitzer} Space Telescope for the Great Observatories 
Origins Deep Survey (GOODS) to detect and study a collection of Lyman-break
galaxies at $z\approx 6$ to 5 in the Hubble Ultradeep Field (HUDF), six of 
which have spectroscopic confirmation.   At these redshifts, IRAC samples 
rest-frame optical light in the range 0.5 to 0.8$\mu$m, where the effects 
of dust extinction are smaller and the sensitivity to light from evolved 
stars is greater than at shorter, rest-frame ultraviolet wavelengths 
observable from the ground or with the Hubble Space Telescope.  As such, 
it provides useful constraints on the ages and masses of these galaxies' 
stellar populations.  We find that the spectral energy distributions for 
many of these galaxies are best fitted by models of stellar populations 
with masses of a few~$\times 10^{10} M_\odot$, and with ages of a few 
hundred million years, values quite similar to those derived for typical 
Lyman break galaxies at $z \approx 3$.  When the universe was only 1 Gyr 
old, some galaxies had already formed a mass of stars approaching that 
of the present--day Milky Way, and that they started forming those stars 
at $z > 7$, and in some cases much earlier.  We find that the lower limits 
to the space density for galaxies in this mass range are consistent with 
predictions from recent hydrodynamic simulations of structure formation 
in a $\Lambda$CDM universe.   All objects in our samples are consistent 
with having solar metallicity, suggesting that they might have already been 
significantly polluted by metals and thus are not comprised of ``first stars''. 
The values for dust reddening derived from the model fitting are low or zero, 
and we find that some of the galaxies have rest-frame UV colors that are even 
bluer than those predicted by the stellar population models to which we 
compare them.  These colors might be attributed to the presence of very 
massive stars ($> 100 M_\odot$), or by weaker intergalactic HI absorption 
than what is commonly assumed.

\end{abstract}


\section{Introduction}

   A striking result from studies of galaxy formation and evolution is that
galaxies as massive as $\sim 10^{11}$--$10^{12}M_\odot$ existed at $z\approx 1$
or even out to $z\approx 2$ (e.g., Fontana et al. 2004, Cimatti et al. 2004,
Glazebrook et al. 2004, Daddi et al. 2005), and galaxies with masses 
$\sim 10^{10}M_\odot$ are already rather commonly seen at $z\approx 3$
(e.g., Papovich, Dickinson \& Ferguson 2001; Shapley et al. 2001). Many such
massive galaxies apparently have well-evolved stellar populations, {\it i.e.},
they must have formed their stars well before the epoch at which they are
observed. For example, about 20\% of the LBGs in the $z\approx 3$ sample of
Shapley et al. (2001) have stellar masses $\geq 10^{10}M_\odot$ and inferred 
ages $>1$Gyr, implying formation redshifts of $z_f>5$. Infrared selected samples
have identified candidates for more massive, evolved galaxies at $z \approx 2$--3
(Franx et al.\ 2003; Yan et al.\ 2004; Labb\'e et al.\ 2005), with inferred 
stellar masses that can exceed $10^{11} M_\odot$.  The dominant luminosity
components for some of these objects are best explained by an old stellar
population with ages of 1.5--2.5 Gyr, which, taken at face value, suggests that
they formed no later than $z_f = 5$, and possibly as early as
$z_f = 15$--20.

   If the inferences from these studies are correct, we should see massive
galaxies at $z\approx 5$ and beyond. One of the central science drivers of the
Great Observatories Origins Deep Survey (GOODS) {\it Spitzer} Legacy Science
Program is to constrain the stellar masses of $L^*$ galaxies out to 
$z\approx 5$ with the rest-frame near-IR observations made by the Infrared
Array Camera (IRAC; Fazio et al. 2004). 
In fact, the performance of this instrument in its 3.6 and 4.5$\mu$m channels
exceeds the pre-launch expectations because the in-flight point spread 
function (PSF) exceeds the telescope specifications.
This enables the detection of galaxies as distant as $z \approx 6$.
At these redshifts, the observed optical and near-infrared data sample 
rest-frame ultraviolet light, which is sensitive to young star formation and 
dust absorption, but which gives relatively little information about the age or
mass of a galaxy's stellar population. {\it Spitzer} and IRAC offer access to
the rest-frame optical light which is less subject to dust extinction and more
sensitive to the longer-lived stars that dominate the stellar mass. Egami et 
al.\ (2005) have used IRAC to detect a galaxy at $z \approx 6.7$ lensed by a 
foreground cluster, while Eyles et al.\ (2005) have studied two galaxies
at $z \approx 6$ that are detected in the GOODS IRAC data. In both papers, the
spectral energy distributions were used to estimate the stellar masses of these 
galaxies, and to demonstrate the presence of stars formed several tens to
several hundred million years earlier.

   In this paper, we study the rest-frame UV to optical photometry of
$z\approx 6$ and 5 (hereafter ``high-$z$") galaxies selected from the
Hubble Ultra Deep Field (HUDF\footnote{See http://www.stsci.edu/hst/udf}; PI.
S. Beckwith), using the HST ACS/NICMOS data and the first epoch of GOODS IRAC 
observations of the {\it Chandra} Deep Field South (CDF-S). While the GOODS
IRAC data cover a much wider area, an important advantage of analyzing galaxies
in the HUDF is the availability of very deep near-infrared imaging, in two
bands, from HST/NICMOS (Thompson et al.\ 2005) for a significant number of
objects. Without such data, there is rather little constraint on the rest-frame
ultraviolet properties of galaxies at $z \gtrsim 5$ compared to what other
studies have had available when analyzing Lyman-break galaxies at lower
redshifts. The UV luminosity of a galaxy is an important constraint on its rate
of on-going star formation, while the UV spectral slope is sensitive to dust
reddening. When fitting population synthesis models to galaxy photometry, it is
important to have constraints on both of these quantities in order to derive
reliable star formation rates, extinction, stellar population ages, and 
mass-to-light ratios. At $z \approx 6$, flux in the ACS $i_{775}$ band, and to
a lesser extent also the $z_{850}$--band (depending on the galaxy redshift), is
suppressed by the stochastic effects of the Lyman~$\alpha$ forest. The ACS 
bands alone are insufficient for reliably measuring the UV luminosity of 
$z \approx 6$ galaxies, and the UV slope cannot be constrained without redder
bandpasses. At $z \approx 5$, the $z_{850}$ band gives a reliable luminosity
at $\sim 1500$\AA, but the spectral slope is again unconstrained without 
near-infrared data.  These high--redshift galaxies are extremely faint, and
only the brightest are detected even in the very deep near-infrared data 
available for the GOODS project from ISAAC on the VLT, and even then at fairly
low signal-to-noise ratios. The NICMOS HUDF data are much deeper and detect
all galaxies from our sample that fall within their field of view, providing
two more photometric bands that cleanly constrain the UV luminosity and
spectral slope.

    The paper is organized as follows. The high-$z$
galaxy samples are briefly described in \S 2, and the IRAC counterparts of
these galaxies are discussed in \S 3. The stellar population synthesis models
that we use to fit the observed spectral energy distributions (SEDs) of these
galaxies are described in \S 4.
The constraints on the stellar populations at $z\approx 6$ and 5 are given in
\S 5 and 6, respectively. We briefly compare our results with recent
hydrodynamic simulations in \S 7, and conclude the paper with a summary in
\S 8.  We denote the F435W, F606W, F775W, and F850LP bands of ACS
as $B_{435}$, $V_{606}$, $i_{775}$, and $z_{850}$, respectively, and the F110W
and F160W bands of NICMOS as $J_{110}$ and $H_{160}$, respectively. In one
occasion we also use the $K_s$-band photometry obtained by the ISAAC instrument
at the VLT. All magnitudes are in the AB system. 
Throughout this paper, we adopt the following cosmological parameters based
on the Wilkinson Microwave Anisotropy Probe result (WMAP) from Spergel et al.
(2003):
$\Omega_M=0.27$, $\Omega_\Lambda=0.73$, and $H_0=71$ km$\,$s$^{-1}\,$Mpc$^{-1}$
($h=0.71$).

\section{Samples of Galaxy Candidates at $z\approx$ 6 and 5 in the HUDF}

    The sample of $z\approx 6$ galaxy candidates used in this paper is from
Yan \& Windhorst (2004; hereafter YW04), which consists of 108 objects to
$z_{850}\leq 30.0$ mag. These candidates
were selected as $i_{775}$-band dropouts using the criteria of 
(1) $(i_{775}-z_{850})\geq 1.3$ mag and (2) non-detection in both the $B_{435}$
and the $V_{606}$ bands ($S/N<2$). These criteria are similar to those used in
Dickinson et al. (2004). The targeted redshift window is 
$5.5\lesssim z \lesssim 6.5$.
There are 6 multiple systems among these candidates (see Table 1 of YW04),
where a multiple system is defined as a group whose members are within $1''$ of
each other. Such a multiple system cannot be resolved by IRAC,
and will show up as a single IRAC source if detected. 

   The sample of $z\approx 5$ galaxy candidates is the $V_{606}$-band
dropout sample of Yan et al. (in preparation). Briefly, the selection
criteria are (1) 
\begin{displaymath}
      \left\{ \begin{array} {ll}
       (V_{606}-i_{775}) > 1.2, & \textrm{if}\,\,\, (i_{775}-z_{850})\leq 0.3\\
       (V_{606}-i_{775}) > 0.4 + 2.67\times (i_{775}-z_{850}), & \textrm{if}\,\,\, 0.3<(i_{775}-z_{850})\leq 0.6 \\
       (V_{606}-i_{775}) \geq 2.0, & \textrm{if}\,\,\, (i_{775}-z_{850}) > 0.6
              \end{array} \right.
\end{displaymath}
and (2) non-detection in the $B_{435}$ band ($S/N<2$). These criteria are
similar to
those of Giavalisco et al. (2004b), but are fine-tuned to better suit the
high S/N HUDF data, which have a tighter low-redshift galaxy locus in the
color space. The targeted redshift window is
$4.5\lesssim z \lesssim 5.5$. This sample consists of $\sim 550$ candidates,
among which 95 objects form 43 multiple systems. 

\section{IRAC Counterparts to High-$z$ Galaxy Candidates}

    The IRAC data used in this paper are the mosaics of the first epoch of 
GOODS observations of the CDF-S
\footnote{See http://data.spitzer.caltech.edu/popular/goods/20041027\_enhanced\_v1.}. 
These data are discussed in detail in Dickinson et al. (in preparation), and
are also briefly described in Yan et al. (2004). To summarize, the middle
one-third of the GOODS CDF-S field, which includes the HUDF, has been observed
in all four IRAC channels with a nominal exposure time of $\sim$ 23.18 hours per
pixel. The final drizzle-combined mosaics have a pixel scale of $0.6''$, or
approximately half of the native IRAC pixel size. These IRAC mosaics are
registered to the same astrometric grid as the GOODS ACS images of
the same area, which is also the astrometric frame of the ACS HUDF.
Sources were detected in a weighted sum of the $3.6\mu$m and 4.5$\mu$m images.
The IRAC photometric catalogs used here are somewhat different from those
used in Yan et al. (2004) in the sense that the source extraction parameters
for the current catalogs are better tuned for deblending objects. 

\subsection{Matching high-$z$ objects in the IRAC images}

  As the FWHM of the PSF in the 3.6$\mu$m
channel is about 1.8$''$, the matching of the high-$z$ galaxy candidates and
the IRAC sources was done with a generous, 2$''$ search radius. The matched
sources were then visually inspected to ensure that the identifications were
secure. While the rate of the mis-identification depends on a number of
factors (such as the brightness of both the source and its neighbours, and
their proximity), we found that roughly one-forth of
the identifications in our sample matched unrelated objects. These false
identifications were thus rejected. Furthermore, we limit our study only to
the most reliable identifications. In order to avoid
ambiguity in interpreting the measured fluxes, we do not include any sources
that are blended with foreground objects. Neither do we include 
any sources that are fainter than $m_{3.6\mu m}=26.4$ mag (the formal
$S/N=5$ limit for isolated point sources in the 3.6$\mu m$ channel).
In practice, all but two of the matched objects have $m_{3.6\mu m} \leq 25.3$
mag.

   Because of the precise alignment of
the IRAC mosaics to the GOODS ACS images, the securely matched IRAC sources 
always have centroids within 0.6$''$ from the centroids as measured in the ACS
images\footnote{The astrometric accuracy, which is based on brighter point
sources, is always good to $0.2^{''}$ in all channels.}. In total, seven
$z\approx 6$ objects (in 3 multiple systems) and twenty-two $z\approx 5$
candidates (3 single objects and 8 multiple systems) were securely matched with
IRAC sources\footnote{While there seems to be a larger fraction of multiple
systems than single objects, we have not yet found sufficient evidence that
IRAC preferably identified multiple systems.}. Since none of the multiple
systems is resolved by the IRAC images, from now on we will not distinguish
whether an IRAC source is a multiple system or a single object, and will simply
call it an ``object"\footnote{For these multiple systems, we emphasize
that they satisfy our high-$z$ color selection criteria either when counting
their components individually or when combining the components as a whole.}.
Therefore, our IRAC-detected high-$z$ sample consists of
three $z\approx 6$ objects and eleven $z\approx 5$ objects. Among them, two of
the $z\approx 6$ galaxies and three of the $z \approx 5$ galaxies have been
spectroscopically confirmed (Dickinson et al. 2004; Stanway et al. 2004;
Malhotra et al.\ 2005; Stern et al., in preparation; see \S3.4).
All the objects in our sample are clearly
detected in 3.6$\mu m$, and some of them are detected in 4.5$\mu m$ as well,
where the data are somewhat shallower. The background noise is substantially
larger and the PSF broader in the 5.8$\mu$m and 8.0$\mu$m IRAC channels, and
none of the galaxies is significant detected at those wavelengths.
ACS and IRAC
image cutouts (and, where available, NICMOS as well) are displayed in Fig. 1
and 2 for the the $z\approx 6$ and $z\approx 5$ galaxies, respectively.

   Perhaps unsurprisingly, these IRAC-detected objects are also some of the 
optically brightest galaxies in the ACS high-$z$ sample. The faintest 
$z_{850}$-band
magnitude of the three $z\approx 6$ sources is 26.88 mag (\#7ab after adding
the fluxes from its two members; detailed in \S 3.2), while the faintest
$i_{775}$-band magnitude of the eleven $z\approx 5$ sources is 26.74 mag
(\#48ab after combining its two members). 
Six objects with $z_{850} < 26.9$ in the $z\approx 6$ sample, and 19 objects
with $i_{775} < 26.9$ in the $z \approx 5$ sample, are not included in the
present analysis.  Only the faintest $z\approx 6$ object and the three faintest
$z\approx 5$ objects are certainly not detected in the 3.6$\mu$m IRAC data at
the 2$\sigma$ level. The others are not included because they are blended with
other, nearby IRAC sources and were thus rejected during visual inspection
regardless of whether they were detected.

\subsection{Photometry}

    When studying SEDs measured by different instruments at widely separated
wavelength ranges, one of the largest challenges is to measure a consistent
fraction of the light from each object in each passband such that the spectral
energy distribution is reliable. In our case, we utilize matched-aperture
photometry as much as possible (using SExtractor of Bertin \& Arnouts 1996),
and choose to bring the measurements as close to the total magnitudes as
possible. Our current best-effort photometry of these IRAC-detected 
$z\approx 6$ and 5 galaxy candidates is given in Table 1 and 2, respectively. 

  For the $z\approx 6$ candidates, the ACS magnitudes ($i_{775}$ and $z_{850}$)
are taken from YW04, which are SExtractor ``MAG\_AUTO''
magnitudes measured through matched apertures using the $z_{850}$ mosaic as the
detection image. The ACS magnitudes ($V_{606}$, $i_{775}$ and $z_{850}$) of the
$z\approx 5$ candidates are also ``MAG\_AUTO'' values, but were obtained using
the $i_{775}$ mosaic as the detection image. For the multiple systems (which
are not resolved by either the NICMOS images or the IRAC images), their 
combined magnitudes are listed as well. 

   For those sources that are detected in the NICMOS HUDF images, their $J_{110}$
and $H_{160}$ magnitudes are based on the NICMOS vs. ACS color indices measured
through matched apertures using re-binned ACS mosaics as the detection
images. The high-resolution ($0.03''$/pixel) ACS mosaics were first 
block-averaged by $3\times 3$ to the resolution of the NICMOS mosaics 
($0.09''$/pixel), and then registered to the NICMOS mosaics\footnote{We used
the ``rotated" (North-up and East-left) NICMOS images included in the version
1.0 public data release made by the NICMOS HUDF team on March 9, 2004.}.
SExtractor was run using the degraded $z_{850}$ image and $i_{775}$ image as
the input detection images to extract the NICMOS vs. ACS colors for the
$z\approx 6$ and $z\approx 5$ objects, respectively. These color indices were
then added to the $z_{850}$ ($i_{775}$) magnitudes to obtain the final NICMOS
magnitudes listed in Table 1 (2).

   Because the angular resolution of the IRAC images is approximately 
10-20$\times$ broader than that of the HST NICMOS and ACS images, simple 
rebinning and matched aperture photometry is not applicable, and a different
method must be used. The GOODS IRAC images are very deep, and crowding by 
neighboring objects can be a problem for photometry, even for objects selected
to be relatively isolated, like our high-$z$ sample. We minimize this by using
small apertures (3\arcsec\ diameter) and a correction to total flux. The
aperture corrections were derived from extensive Monte Carlo simulations in
which artificial galaxies with varying magnitudes and sizes were inserted into
the images and then detected and measured using the same SExtractor cataloging
parameters used for the real data. Because our objects are compact, we apply
aperture corrections derived for simulated galaxies with half-light radii 
$< 0\farcs5$, which average -0.55 mag and -0.60 mag for the 3.6$\mu$m and 
4.5$\mu$m channels, respectively. The corrected aperture magnitudes for IRAC
are given in Tables 1 and 2.

   In addition to the ACS, NICMOS and IRAC data, we also use the deep
VLT/ISAAC $K_s$-band images (Vandame et al., in preparation; Giavalisco et al.
2004a). Only one object, \#6ab in the $z\approx 5$ sample, is significantly 
($S/N>3$) detected. While the inclusion or exclusion of this data point is not
significant for this object in particular, we use it for the sake of completeness.

\subsection{Assessment of systematic errors in photometry}

   The systematic errors in IRAC photometry are largely
caused by the source blending. To estimate such systematic errors, we used two
more methods to estimate the IRAC magnitudes of these sources. While none of
these methods (including the aperture photometry) completely eliminates the
contamination caused by blending, they are affected differently. By comparing
these three different kinds of magnitudes, we can estimate the effect of 
contamination to these sources.

   One of the additional estimates is the usual ``MAG\_AUTO'' magnitude 
measured by SExtractor. The other one is based on point spread function (PSF)
fitting, which is justified by the fact that these sources are very nearly
point sources at IRAC's resolution. The centroids of the sources measured from
the ACS images are used as priors to the fitting algorithm. The total fitted
magnitude of a multiple system is based on the sum of the fitted fluxes of the
individual components. We use the root-mean-square (rms) dispersion of the
three types of magnitudes as a measure of the systematic plus random 
measurement errors in these crowded images. The zeropoint calibrations for IRAC
are believed to be accurate at the 10\% level or better, but we allow for
additional systematic uncertainty in the absolute photometry at this level,
and therefore add (by simple summation because it is a systematic term) an 
additional 0.1 mag to obtain the final error estimates for these objects
(Tables 1 and 2).

   For the ACS and the NICMOS photometry, we assess the possible systematic
errors by comparing our results against those of others. We use the catalogs
included in the HUDF ACS and NICMOS data release (hereafter ``the public
catalogs'') as well as other published results in the literature for this
comparison. We put the emphasis on the comparison of colors because
they play a more important role than anything else in the SED analysis. 
Given the different source extraction procedures in the different studies,
colors are also the most robust metric for such a comparison. 

    Since the public ACS catalog is largely an $i_{775}$-band-based catalog
({\it i.e.}, the $i_{775}$-band image was used to detect sources and to define
source extraction apertures), we only compare the photometry of the
$z\approx 5$ sources. Object \#10 and \#14ab are excluded from this comparison
because they are not in the public ACS catalog.

   For the majority of these sources, both the $(V_{606}-i_{775})$ and 
$(i_{775}-z_{850})$ indices agree to within 0.1 mag in these two catalogs.
Given that the two catalogs use different flavors of magnitude (the ACS public
catalog uses SExtractor ``MAG\_ISO'' while ours uses ``MAG\_AUTO''), and that
most of the sources have irregular morphologies, such an agreement is
reasonable, and the size of the discrepancies reflects the impact of choosing
different photometric apertures. 

    There are three sources whose $(V_{606}-i_{775})$ colors differ by $>0.1$
mag, namely, \#4ab, 15ab and 48ab. Such 
discrepancies are mainly caused by the large photometric errors in the
$V_{606}$-band, and can be explained by the random errors ({\it i.e.}, those
derived based on the $S/N$ of the sources) associated with these colors.
Only \#15ab has $(i_{775}-z_{850})$ different by $>0.1$ mag. Our measurement
gives an index of $0.13\pm 0.04$ mag, while the ACS public catalog gives 
0.28$\pm$0.03 mag. This very irregular object has also been studied in detail
by Rhoads et al. (2005), who used different isophotal apertures on a slightly
different version of the HUDF images. Those authors give an even more different
value of $(i_{775}-z_{850})=0.47\pm 0.02$ mag (after combining their ``core''
and ``plume'' components). We believe that the discrepancies of different
results can also be attributed to the different choices of photometric
apertures and the contamination from the two physically unrelated close
neighbors.

    The means of the color index differences are $0.01\pm 0.12$ mag and 
$0.02\pm 0.08$ mag for $(V_{606}-i_{775})$ and $(i_{775}-z_{850})$, 
respectively. The useful quantities are the dispersions, which tell us
how these colors can be different because of the systematic errors in their
photometry. Given the $V_{606}$-dropout nature of these sources, it is
reasonable to assume that the errors in both $i_{775}$ and $z_{850}$ are
comparable, while those in $V_{606}$ are larger. Therefore, we assign these
sources additional errors of 0.056 mag to both $i_{775}$ and $z_{850}$, and 
0.106 mag to $V_{606}$, respectively. These numbers, while added in quadrature,
give the corresponding values of dispersion in the relevant colors.
For the $z\approx 6$ objects, we assume that their $(i_{775}-z_{850})$
colors suffer a comparable amount of systematic error as the 
$(V_{606}-i_{775})$ colors of the $z\approx 5$ sources. Thus we assign them
additional errors of 0.106 mag to $i_{775}$, and 0.056 mag to $z_{850}$,
respectively. The errors reported in Table 1 and 2 are the sum of such 
additional systematic error estimates and the random errors based on their
$S/N$.

   We compare our $(i_{775}-z_{850})$ colors of the three $z\approx 6$
sources to the values (based on a fixed-size, 0.5$^{''}$-diameter aperture)
reported in Bunker et al. (2004), and find that the differences can all be 
explained by the systematic uncertainties that we adopt.

   Since our NICMOS magnitudes are derived based on matched-aperture
photometry between ACS and NICMOS, we believe those magnitudes are superior
to any others for the purpose of SED analysis. As these magnitudes are
calculated by adding NICMOS vs. ACS colors to the ACS magnitudes ({\it i.e.},
adding $J_{110}-i_{775}$ or $H_{160}-i_{775}$ to $i_{775}$ for $z\approx 5$
objects, and adding $J_{110}-z_{850}$ or $H_{160}-z_{850}$ to $z_{850}$ for
$z\approx 6$ objects), these values should ``inherit'' the systematic errors of
the ACS magnitudes. Therefore, we assign an additional error of 0.056 mag to
both bands. Compared to our photometry, the ``MAG\_AUTO'' values in the NICMOS
public catalog (version 1.0) on average are fainter by 0.12 mag in $J_{110}$ 
and 0.24 mag in $H_{160}$, respectively. The NICMOS photometry of the three
$z\approx 6$ objects has also been discussed by Stanway et al. (2004), and
their values agree to ours to within the quoted errors (0.1--0.2 mag). Rhoads
et al. (2005) also discussed the NICMOS colors of object \#15ab in our 
$z\approx 5$ sample; while their $J_{110}-H_{160}$ colors are bluer by 0.19 mag
as compared to our measurements, such a difference can still be understood if
we consider the quoted errors.

\subsection{Galaxies with confirmed redshifts}

   Two objects in the $z\approx 6$ sample have spectroscopic redshift
measurements. Object \#1 was observed with the Keck
and VLT observatories and with the ACS grism, and has $z=5.829$ (Dickinson
et al. 2004; Stanway et al. 2004; Vanzella et al. 2005). Based on its
morphology, the secondary component \#1b seems to be a ``filament" extending
from \#1a (see Fig. 1), and therefore it is taken as at the same redshift as
\#1a. 
   
   The Grism ACS Program for Extragalactic Science (GRAPES; Pirzkal et al.
2004a) has confirmed the high-$z$ nature of
\#5abc (Malhotra et al. 2005). However, the components \#5ab (not resolved
by the GRAPES; ID 3450) and \#5c (ID 3503) are identified at rather different
redshifts of $5.9$ and $6.4$, respectively. As their contributions to the IRAC
fluxes cannot be separated, as a first-order approximation we treat them as at
the same, but not-yet-determined redshift. We will return to the discussion of
the redshift of this system in \S 5.1 again.

   Four objects in the $z\approx 5$ sample have spectroscopic redshifts. Object
\#6ab, 7ab and 15ab have been observed at Keck and are confirmed to be at
$z=4.65$, 4.78 and 5.49, respectively (Stern, in preparation). The GRAPES
has also measured $z = 4.88$ and 5.52 for \#7ab and 15ab, respectively
(Xu et al. 2005; see also Rhoads et al. 2005 for an initial identification
of \#15ab, which gave $z=5.42$ for this source). Here we adopt the values
obtained at Keck. The GRAPES has also measured $z = 5.05$ for object 
\#4ab (Xu et al. 2005), which we adopt in our discussion. 

\section{Overview of Stellar Population Synthesis Models and Procedures}

   We use the stellar population synthesis models of Bruzual \& Charlot (2003;
BC03) to analyze the SEDs of these IRAC-detected high-$z$ galaxies. For ease
of comparison with other studies, we use a Salpeter (1955) initial mass
function (IMF). This IMF is incorporated in BC03 models with lower and upper
mass cut-offs at 0.1 and 100 $M_\odot$, respectively. Recent work has favored
a flatter IMF below 1 $M_\odot$ (e.g., Chabrier, 2003).  For our purposes, the
shape of the IMF at low masses only affects the derived stellar mass-to-light
ratios ($M/L$) in a way that is nearly independent of stellar population age.
The Chabrier IMF yields masses that are 0.53--0.60 times those derived using
a Salpeter IMF; the same will be true for galaxies at lower redshifts to which
we might compare our results for the high-redshift population. Instead, changing
the slope or shape of the IMF at high masses can have more pronounced effects on
galaxy colors and their time evolution, and on the relative mass-to-light
ratio as a function of stellar population age. E.g., a top-heavy IMF will produce
more light and have smaller $M/L$ at younger ages, and the luminosity fade more
quickly with time.  As yet, we have little or no observational constraint on the
IMF for galaxies at $z \approx 5$ to 6, and we therefore adopt the Salpeter form
to simplify comparison with other published work, but systematic uncertainties
related to the choice of the IMF should be kept in mind. We will return briefly
to one IMF-related issue in Sections 5.2 and 6.2.

    The model spectra generated by the BC03 code are shifted across the range of
expected redshifts with a step size $\Delta z = 0.1$, and attenuated by 
intergalactic HI  absorption 
according to the recipe of Madau (1995). The model spectra are then integrated
through the bandpass response curves and compared to the observations.
We consider BC03 models with five metallicities $Z$, from $Z_\odot/200$ to
solar. We explore a variety of star formation histories (SFHs): instantaneous
bursts (or simple stellar populations, SSPs); continuous but exponentially 
declining star formation with e-folding timescales from $\tau = 10$~Myr to 1
Gyr; and continuous star formation at a constant rate.
We also consider dust reddening following the empirical law of
Calzetti et al. (2000) within the range of $E(B-V)=0$--2.3 mag.
An important constraint on the models is that the inferred ages 
should not exceed the age of the universe at the corresponding redshifts. For
example, if a set of models give photometric redshift of $z_p=6.0$ for an 
object, the legitimate models should only be those that have ages less than
0.95 Gyr (the age of the universe at $z=6$ in our adopted cosmology). 

    In some cases, the observed photometry cannot be satisfactorily fit by a
single-component model of the sorts that we have considered. One reason
why this may occur is because the real SFH is likely more complicated than
these basic models. Therefore, we consider the superposition of two models when a
single model fails to explain the observed SED. This ``two-component" approach
has been used by Yan et al. (2004) to explain the weak optical fluxes of the
IRAC-selected extremely red objects (IEROs). It is similar to the ``maximum $M/L$''
method of Papovich, Dickinson \&
Ferguson (2001), where they derived the maximum stellar mass allowed by
the observed SED of a given galaxy through breaking its light into the 
contributions from a maximally old component (formed at $z_f=\infty$) and from
a young component. The differences between the two methods are that introducing
one more component in our current approach is driven by what is
necessary ({\it i.e.}, when the one-component models fail) rather than what is
allowed, and that each of the two components can be formed at any arbitrary
$z_f$ as long as it is larger than the redshift at which a given galaxy is
observed. This approach is also similar to that of Berta et al. (2004), but is
different in the sense that they considered an arbitrary large number of
components represent by SSPs, while we limit our method to only two components
(but not necessarily SSPs) such that the major processes can be seen more 
clearly.

   We carry out the fitting in flux density units (instead of magnitudes), and
minimize $\chi^2$ to find the best-fitting model. When only single-component
models are involved, the possible free parameters are:
redshift ($z$), age ($T$), stellar mass ($M$), star formation history ($\tau$;
SSP is treated as $\tau=0$), reddening ($E(B-V)$), and metallicity ($Z$). 
When a spectroscopic redshift is available, this parameter is fixed. When 
two-component models have to be evoked, we require both components be at the
same redshift and have the same metallicity. 

   In the practice of minimizing $\chi^2$, the goodness-of-fit is usually 
measured by the reduced-$\chi^2$, which is defined as $\chi^{'2}=\chi^2/\nu$,
where $\nu$ is the degrees of freedom of the given problem. In our case, $\nu$
is the number of available bandpasses minus the number of free parameters. The
objective of our analysis is not only to find the parameters that best describe
the SEDs ({\it i.e.} the ones giving the smallest $\chi^{'2}$), but also to find
the ranges within which the parameters can still give satisfactory fits to the
SEDs ({\it i.e.}, the ones giving acceptable $\chi^{'2}$). While in principle
the latter can be achieved by drawing constant $\chi^2$ boundaries as confidence
limits on the estimated parameters, in practice it is difficult to take this
approach. The fundamental difficulty is that the number of available
bandpasses is limited, and thus for most of objects in our samples the values
of $\nu$ would be zero or negative if we were to vary all the parameters.

   To overcome this difficulty, we take the following approach. For a given
object, we carry out the fitting procedure in separate runs, and require 
$\nu\geq 1$ in each run. This is achieved by ``freezing" different parameters in
different runs separately. Whenever possible, we simultaneously fit for the
maximum number of free parameters allowed, {\it i.e.}, making $\nu=1$. For
example, for an object that has five bands available, we can fit for four free
parameters at the same time if using single-component models. A possible run 
then can fit for ($T, M, z, Z$) with ($\tau$, $E(B-V)$) fixed, or 
($T, M, z, \tau$) with ($Z$, $E(B-V)$) fixed, and so on. 
We start from an arbitrary combination of maximum
number of free parameters (but always including $T$ and $M$, and also $z$ if it
is unknown), and find the best solution in this run. Of course, this requires we
fix other parameters to some initial values based on our best-guess. We deem
a fit to be of high quality if it has $\chi^{'2}\leq 1+\sqrt{2/\nu}$
(Press et al. 1992). When $\nu=1$, which is true for most cases, this criterion is
$\chi^{'2}\leq 2.41$. If the best-fit in this run is a high-quality solution, 
we then ``freeze'' the 
parameters (anything other than $T$ and $M$) to the values of this best-fit,
and begin finding the best-fit in a different run involving a combination of
different parameters. We repeat this process until
all free parameters are exhausted, and then declare the best solution in the
last run as the best-fit model to the object under question. Finally, we alter
all the parameters around the best-fit values and see to what extent we can
still obtain high-quality fits. This approach has a number of 
caveats, one being that such a solution might be a local best-fit but not a 
global best solution. Here we have to neglect such possible caveats.

\section{Stellar populations at $z\approx 6$}

   Here we discuss in detail the fitting results for the three objects in
the $z\approx 6$ sample. The passbands that go into the fitting procedure
are from $i_{775}$ to 3.6$\mu$m (see Table 1). The best-fit models are shown
in Fig. 3, and their parameters are summarized in Tab. 3.

\subsection{Evidence of evolved, massive systems}
  
   We discuss the three $z\approx 6$ objects in order of easiness in their
fitting.

   {\bf Object \#7ab:} This object can be fitted by single-component models.
The best-fit model ($\chi^{'^2}=0.66$) to this object is a short burst
($\tau=10$ Myr) at $z_p=5.9$, with solar metallicity and no reddening. The
derived age is 0.1 Gyr and the stellar mass is $4.7\times 10^9 M_\odot$. 
Around this best-fit, we find following constraints on the free parameters:

  (1) $\tau$: At $0\leq \tau\leq 10$ Myr (including SSP), we can find 
high-quality solutions of the same quality ($\chi^{'2}<0.7$) as the best-fit,
and the derived age
and mass are also effectively the same. The quality of fit is rapidly reduced
at $\tau >10$ Myr, although it does not seem to depend on $\tau$ in a linear
fashion. We are able to find high-quality solutions sporadically with 
increasing $\tau$ all the way to $\tau\leq 70$ Myr, and if introducing
$E(B-V)\approx 0.2$ mag such solutions can still be found until $\tau\leq 0.1$
Gyr. Such solutions with $\tau >10$ Myr, however, seem unstable, as changing
their ages ($T$) by only one step in either direction will immediately make 
$\chi^{'2}>5$.

  (2) $Z$: While lowering metallicities always (although non-linearly) gives at
least two-times worse $\chi^{'2}$, we cannot reject sub-solar
metallicties at a significant level. Such models generally produce an older age
and a larger mass. 

  (3) $E(B-V)$: The quality of fit becomes significantly worse if $E(B-V)>0$.
At $E(B-V)=0.23$ mag, any solution has $\chi^{'2}$ at least twice as bad as the
best fit. If $E(B-V)\geq 0.45$ mag, no high-quality solution can be found. 

  (4) $T$ \& $M$: We find high-quality solutions only within 
$0.05 \leq T \leq 0.1$ Gyr and 
$1.4\times 10^9 < M < 1.8\times 10^{10} M_\odot$.

   We do not find any acceptable single-component models to fit object \#1ab
and 5abc. The best-fit single-component models to \#1ab and 5abc have
$\chi^{'2}=10.49$ and 2.78, respectively, and none of them satisfy the 
high-quality criterion ($\chi^{'2}\leq 2.41$ in these cases) that we adopted.
For the sake of completeness, we list the parameters of these fits:
($T=0.9$ Gyr, $\tau=0.9$ Gyr, $M=1.9\times 10^{10}M_\odot$, $E(B-V)=0$, 
$Z=0.2Z_\odot$) for \#1ab, and 
($T=0.9$ Gyr, $\tau=0.5$ Gyr, $M=1.6\times 10^{10}M_\odot$, $E(B-V)=0$,
$Z=0.2Z_\odot$) for \#5abc.
Note that both models have a very old age of 0.90 Gyr (the age
of the universe at $z=5.8$ and $5.9$ is 0.99 and 0.97 Gyr, respectively),
which means that one has to push the single-component models almost to their
limits in order to obtain even these mediocre fits.

    Therefore, we have to consider
two-component models for these two objects. From here on, we use subscripts
``E'' (evolved) and ``Y'' (young) to denote the parameters for the primary and
secondary components, respectively. As we require both components have the
same redshift and metallicity (see \S 4), no subscript is put to these two
parameters.

   {\bf Object \#5abc:} The best-fit ($\chi^{'2}=0.23$) to this
object has $z_p=$5.9 and $Z=Z_\odot$, and
($T_E=0.9$ Gyr, $\tau_E=0.2$ Gyr, $M_E=3.8\times 10^{10}M_\odot$, $E(B-V)_E=0$)
and ($T_Y=0$, $M_Y=1.0\times 10^8M_\odot$,$E(B-V)_Y=0$), respectively. 
We note that its photometric redshift is consistent with the
redshift given by the ACS grism (Malhotra et al. 2005; $z=5.9$) for the
component \#5ab. Excluding the contribution of \#5c (see \S 3.3) from the SED
does not change any of our conclusions.

   (I) High-quality solutions can only be found when the young component
has $T_Y\leq 30$ Myr and $E(B-V)_Y=0$. Its $\tau_Y$, however, is arbitrary,
because any SFH gives effectively the same solution at such a young age.
For the same reason, this component does not pose any constraint on the
metallicity.

   (II) If we assume that both components have the same reddening, {\it i.e.},
$E(B-V)_E=0$, we have following constraints on the primary component:

    (1) $\tau$ \& $T$: High-quality solutions are available only when 
$\tau_E\leq 0.4$ Gyr and $T_E\geq 0.4$ Gyr. Thus the age implies a formation
redshift $z_f\gtrsim 8.9$. The youngest $T_E$ (0.4 Gyr) is found only when 
$\tau_E\leq$ 80 Myr. If $0.2 \lesssim \tau_E \lesssim 0.4$ Gyr, the youngest
$T_E$ will be 0.7 Gyr, implying $z_f\gtrsim 16$. 

    (2) $M$: We find high-quality solutions only when
$1.3\times 10^{10} \lesssim M_E \lesssim 6.2\times 10^{10} M_\odot$.

    (3) $Z$: At one-step lower metallicity, $Z=0.4 Z_\odot$, solutions of
comparable high quality are available. Decreasing metallicity beyond this
value steadily decreases the quality of fit, and no high-quality solutions can 
be obtained if $Z=1/200 Z_\odot$ (the lowest metallicity available in BC03).

   (III) If non-zero $E(B-V)_E$ is allowed ({\it i.e.}, the two components can
have different reddening), most of the parameters for the primary component are
no longer well constrained. Even in this case, however, the stellar mass is
$M>2.0\times 10^{10} M_\odot$ for high-quality solutions. We note that none of
the solutions with non-zero $E(B-V)_E$ has as small $\chi^{'2}$ as the best-fit
model does. If we confine the solutions to solar metallicity,
no high-quality solution can be found at $E(B-V)_E\gtrsim 1.1$ mag. This 
constraint is largely set by the slope of the SED from $3.6\mu$m to $4.5\mu$m
(larger reddening values would require a much steeper slope). We find that most
of the high-quality solutions have $E(B-V)_E\lesssim 0.45$ mag if $Z=Z_\odot$,
which might indicate that the true reddening value is indeed insignificant. For
low reddening values, the high-quality solutions quickly converge to those
obtained with $E(B-V)_E=0$. 

   {\bf Object \#1ab ($z=5.83$):\footnote{This object is one of the objects
discussed by Eyles et al. (2005). Our main results broadly agree with theirs
 in the sense that the best-fit stellar mass is $>10^{10}M_\odot$ and the age
is a few hundred Myrs.}} We find no satisfactory one- or two-component model 
fit to this object if we adopt the normal standard of high-quality. As we
will discuss in detail in the next section, we find that the large $\chi^{'2}$
values are mainly caused by its abnormal $(z_{850}-J_{110})$ color as compared
to the BC03 models. This suggests that the models used for the secondary
(young) component might have some intrinsic shortcomings. As the primary
(evolved) component is more important to our conclusions, and the models do not
have such a
problem in explaining this component, we still use these models in interpreting
this object, but loosen our criterion and compare the fits in a relative sense.
Therefore, we discuss all solutions that have give $\chi^{'2}<10$. We
find a best-fit of $\chi^{'2}=5.6$, which gives 
($T_E=0.5$ Gyr, $\tau_E=0$, $M_E=3.4\times 10^{10}M_\odot$, $E(B-V)_E=0$) and
($T_Y=0.001$ Gyr, $M_Y=2.3\times 10^8M_\odot$, $E(B-V)_Y=0$), respectively.
Both components have solar metallicity. 

    (I) Solutions of $\chi^{'2}<10$ can only be found when the young
component has $T_Y\leq 30$ Myr and $E(B-V)_Y=0$. Similar to the case of 
\#5abc, we have no constraint on the SFH of this component.

    (II) If we require the primary component have the same reddening as the
secondary one, {\it i.e.}, $E(B-V)_E=0$, we have following constraints on the
primary component:

    (1) $\tau$ \& $T$: Solutions of $\chi^{'2}<10$ can only be found
when $T_E\geq 0.2$ Gyr and $\tau_E\leq 0.6$ Gyr. The minimum $T_E$ of 0.2
Gyr implies a formation redshift $z_f\gtrsim 7$. If $0.5\leq \tau_E\leq 0.6$
Gyr, solutions of comparable quality can only be found at $T_E\geq 0.8$ Gyr,
which means $z_f\gtrsim 20$. 

    (2) $M$: Solutions of $\chi^{'2}<10$ can only be found within
$1.4\times 10^{10} \leq M_E\leq 5.7\times 10^{10} M_\odot$.

    (3) $Z$: While we see the general trend of worsening fit with decreasing 
metallicity, we are not able to reject models with lower metallicities.
However, we point out that if $Z=1/200 Z_\odot$, the only 
$\chi^{'2}<10$ solutions available are those having very old $T_E$
($\geq 0.8$ Gyr) and very short $\tau_E$ (very close to SSPs).

    (III) Allowing $E(B-V)_E$ to be different from $E(B-V)_Y$ makes the models
largely unconstrained. However, we find that $E(B-V)_E$ cannot be larger
than 0.4 mag (for the same reason as in the case of \#5abc) and $M$ cannot
be less than $2.0\times 10^{10} M_\odot$, or no solution
of $\chi^{'2}<10$ can be found. 

    The SED analysis for these $z\approx 6$ objects, especially \#1ab and
5abc, strongly suggests that systems as massive as $M\sim 10^{10} M_\odot$
already existed when the universe was less than 1 Gyr old, and they probably
started forming their stars several hundred Myr earlier. While we do not have a
strong constraint on their metallicities, they are consistent with solar abundance, 
implying that they might have already been significantly polluted by metals.
Dust reddening to the three objects in our sample seems to be moderate or even
negligible, and the best-fits to their SEDs always have zero reddening.
Dust with grayer attenuation could, however, cause extinction without reddening.
This would also increase the stellar masses derived for the galaxies.

\subsection{Very blue rest-frame UV colors and hint of very massive stars}

    As mentioned above, \#1ab has an abnormally blue $(z_{850}-J_{110})$ color
as compared to the BC03 models\footnote{The discrepancy is not likely caused
by a zeropoint error in $J_{110}$, as the zeropoints in the NICMOS passbands 
are accurate to 0.01--0.02 mag (R. Thompson, private communication).}. Such a
very blue rest-frame UV color is not
unique to this object, however. In fact, most of the $z\approx 6$ candidates in
YW04 that have NICMOS photometry show such a trend in their $(z_{850}-J_{110})$
colors, some of which even have $(z_{850}-J_{110}) \sim -0.5$ mag (see Fig. 1 of
YW04; see also Stanway et al. 2004). The matched-aperture photometry described
in \S 3.3 further confirms this result, and gives more accurate measurements
of the colors. No matter what the specific SFH is, a galaxy at this redshift
should have a rather flat SED (in terms of $f_\nu$ vs. $\lambda$) in the 
rest-frame UV, i.e., its $(z_{850}-J_{110})$ colors should be close to zero.
The NICMOS $J_{110}$ bandpass heavily overlaps that of the ACS $z_{850}$
bandpass, making it quite difficult to produce colors as blue as those observed
here.

   Fig. 4 compares the measured $(z_{850}-J_{110})$ colors of the $z\approx 6$
galaxies that have NICMOS photometry against the expected $(z_{850}-J_{110})$
colors as a function of redshift as derived from a series of solar metallicity
models. The models are drawn from those mentioned in \S 4, and are of four
types of representative SFH: SSP, $\tau=0.1$ Gyr, $\tau=1.0$ Gyr, and constant
star-formation. The ages of the models are 0.001 Gyr, 0.01 to 0.09 Gyr (0.01 
Gyr step-size), and 0.1 to 1.1 Gyr (0.1 Gyr step-size). The youngest track is
at the bottom, and the oldest track is at the top. The three IRAC-detected
sources are shown as filled squares: \#1ab is put at $z=5.83$ without a
redshift error bar, \#5abc is put at $z=6.0$ with an uneven error bar 
from $z=5.9$ to 6.4, and \#7ab is put at $z=6$ with an error bar indicating
its possible redshift range of $5.5\leq z\leq 6.5$. The open squares (all put at
$z=6$ and with error bars from $z=5.5$ to 6.5) are other $z\approx 6$ objects
that have NICMOS photometry.

   While object \#7ab can be explained reasonably well by a population with
age of $\sim$ 0.1 Gyr, the majority of the sources are problematic. Regardless
of the SFH, the bluest color ``boundary" is always defined by the zero-age track
(almost indistinguishable from the 0.001Gyr track in the figure), which cannot
extend to $(z_{850}-J_{110})<-0.2$ mag at any redshift. We emphasize
that object \#1ab, whose redshift is unambiguously known, is not consistent
with any of these models. Such blue colors cannot be attributed to low
metal abundance, as the most metal-poor (1/200 of solar, or $Z=0.0001$) models
of BC03 have
essentially the same bluest color boundary as shown in Fig. 4. A less severe
intervening HI absorption than that of Madau (1995) might possibly explain the
bluer colors at $z\geq 6$ if $(z_{850}-J_{110})>0$ mag, but cannot explain the
$(z_{850}-J_{110})<-0.2$ mag colors at $z<6$. Another possibility would be the
presence of a strong \Lya emission line in the $z_{850}$ band, which could make
the $(z_{850}-J_{110})$ color bluer than the models predict. In order to 
reproduce the observed colors of \#1ab, however, the \Lya line would have to
have rest-frame equivalent width (EW) $>300$\AA. This is not the case for
\#1ab; we examined its spectrum obtained at the Keck (Dickinson et al. 2004),
and found that it only had a moderately strong \Lya line with rest-frame 
EW=18\AA. Neither is it the case for the major component of \#5abc, which do
not show \Lya emission line based on the grism spectra of the GRAPES program. 

   These very blue rest-frame UV colors might suggest that our model templates
do not sufficiently populate the entire parameter space. One limitation of the
BC03 models is that the high-mass end of the adopted Salpeter IMF cuts off at
100 $M_\odot$. If more massive stars are included, such very blue colors could
be explained. For example, a toy galaxy model consists of only the 
hottest stars ($T_{eff}\sim 30,000 K$; solar metallicity) from Lejeune et al.
(1997) can easily explain the observed $(z_{850}-J_{110})$ colors. The lifetime
of such very hot (thus very massive) stars, however, is very short (in any case 
$< 1 Myr$), which indicates that we might be watching a special episode when
these early galaxies were actively forming the most massive stars.

   As varying the IMF of the BC03 models is beyond the scope of this paper,
here we still adopt these models as they are, but caution that the SED fitting
based on these models will likely have problems in explaining the rest-frame
UV part of some of the SEDs. In spite of this drawback, one conclusion seems 
inescapable from this analysis is that most galaxies shown in Fig. 3 (including
\#1ab and \#5abc) have young components indicative of very recent ($\leq 30$ 
Myr) star formation. Furthermore, the dust reddening in these young components
is likely minimal, otherwise their intrinsic rest-frame UV colors would be even
bluer and more difficult to explain. 
 
\section{Stellar populations at $z\approx 5$}

  In this section, we discuss in detail the fitting results for the
$z\approx 5$ objects. The passbands that go into the fitting process are from
$V_{606}$ to 4.5$\mu m$. The best-fit models are shown in Fig. 5 and 7, and
their parameters are summarized in Tab. 4.

\subsection {SED fitting results}

    We first consider the five objects that have NICMOS photometry. 

   {\bf Object \#6ab ($z=4.65$):} This object can be well explained by single-component
models. As its redshift is known, with its 8 bands of photometry the degrees
of freedom are $\nu=3$. Thus our criterion of high-quality is
$\chi^{'2}\leq 1.82$. While high-quality solutions are available at
all metallicities, the goodness-of-fit improves with smaller $Z$ values. The
best-fit has $\chi^{'2}=0.45$, and has
($T=1.3$ Gyr, $\tau=0.4$ Gyr, $M=2.8\times 10^{10} M_\odot$, $E(B-V)=0$, 
$Z=1/200 Z_\odot$). We find high-quality solutions only within the following ranges:
$E(B-V)\leq 0.23$ mag, $\tau\geq 0.2$ Gyr, $T > 0.5$ Gyr, and
$1.8\times 10^{10} \leq M\leq 5.1\times 10^{10} M_\odot$. The lower bound of
its age implies $z_f\gtrsim 7.3$.

   {\bf Object \#15ab ($z=5.49$):} Our models (either single-component or two-component)
cannot produce any fit with $\chi^{'2}<40$ for this object because of
its abnormally blue $(i_{750}-z_{850})$ color. As discussed in previous section,
such an abnormal color of this object in particular could be explained by a
less severe HI  absorption in the IGM along the sight-line. If we exclude
the $i_{775}$ band from the fitting process based on the argument that our
knowledge about the IGM HI  absorption is imperfect, we can find high-quality
fits using single-component models. The best-fit has $\chi^{'2}=1.63$,
and gives 
($T=1.0$ Gyr, $\tau=0.6$ Gyr, $M=2.2\times 10^{10} M_\odot$, $E(B-V)=0$, $Z=Z_\odot$).
We find no high-quality solutions unless $E(B-V)=0$. On the other hand, we have
no obvious constraint on its metallicity. High-quality solutions are available
if $0.3 \leq \tau\leq 0.7$ Gyr, $0.8 \leq T\leq 1.0$ Gyr, and
$1.8\times 10^{10}\leq M\leq 3.6\times 10^{10} M_\odot$. The age range implies
that the formation redshift of this object is likely $z_f\gtrsim 15.8$.
Rhoads et al. (2005) considered whether this object could be powered by
an active galactic nucleus (AGN), but did not reach a conclusive answer.
We do not find any compelling evidence in its SED that this object harbors
an AGN.

    {\bf Object \#48ab:} Regardless of the values of other parameters, the 
photometric redshift of this object is well constrained at $z_p=5.2$. 
The best single-component fit to this object gives $\chi^{'2}=2.48$,
which does not strictly satisfy our high-quality criterion 
($\chi^{'2}=2.41$ in this case). The parameters of this fit are
($T=0.03$ Gyr, $\tau=0$, $M=9.4\times 10^{8}M_\odot$, $Z=1/5Z_\odot$,
$E(B-V)_E=0$). The large $\chi^2$ of this best-fit comes mostly from the
discrepancy in the IRAC $3.6\mu$m band, where the model is lower than the 
observation by a factor of six. 

Thus we also explore two-component models. We find the following:

    (1) High-quality fits can be obtained only when $E(B-V)_Y=0$. If we limit
the reddening of the primary component also to $E(B-V)_E=0$, the fit only
improves moderately. The best-fit ($\chi^{'2}=1.82$) has a
metallicity of $Z=0.4Z_\odot$, and has 
($T_E=1.0$ Gyr, $\tau_E=0$, $M_E=1.2\times 10^{10} M_\odot$) and
($T_Y=10$ Myr, $\tau_Y=0$, $M_Y=2.2\times 10^{8} M_\odot$). 

    (2) If we allow non-zero $E(B-V)_E$, very high-quality fits
($\chi^{'2}\lesssim 0.3$) can be obtained. However, parameters of these
fits span vastly different ranges and thus are not constrained. For
example, the fitted stellar mass $M_E$ can be arbitrarily high with increasingly
severe reddening. Its age is also arbitrary with different reddening values,
spanning from 10 Myr to 1.0 Gyr.

    Therefore, we consider the fit to this object uncertain at this point.

    The best-fit models of the above three objects are shown in Fig. 5. 
The remaining two objects that have NICMOS photometry, \#4ab ($z=5.05$)
and 7ab ($z=4.78$), cannot be fitted by any models (single- or two-component)
that we considered. The discrepancy between the models and the observations
is mainly in the rest-frame UV, which is very likely a problem intrinsic
to the current BC03 models (for example, the cut-off at the high-mass end of
the adopted IMF) rather than a problem that can be attributed to more 
complicated SFH. 

    Objects without NICMOS photometry can all be well-fitted by
single-component models. As we do not have $J_{110}$ and $H_{160}$ bands that
are critical in determining the rest-frame UV slope and UV-to-optical slope,
the parameters (except their photometric redshifts, which are largely 
determined by their $V_{606}-i_{775}$ colors) are not tightly constrained for
most objects. The details of the fits to these objects are given below. 
Object \#18 is completely unconstrained the model fitting. It has comparatively
red colors in both $V_{606} - i_{775}$ and $i_{775} - z_{850}$, suggesting that
it may lie near the upper end of the redshift range for the $z \approx 5$ color
selection.  With a very blue $(z_{850} - m(3.6\mu m)) = 0.13$, it is difficult
to further constrain the stellar population parameters of the object in the 
absence of additional near-infrared photometry. While this object is not further
discussed here, its SED is shown in Fig. 8 for completeness.

   {\bf Object \#10:} The best-fit ($\widetilde\chi^2=0.51$) to this object has 
($z_p=5.2$, $T=0.4$ Gyr, $\tau=0.2$ Gyr, $M=9.8\times 10^{9} M_\odot$,
$E(B-V)=0$, $Z=Z_\odot$). High-quality solutions can be found if
($E(B-V)\leq 0.23$ mag, $T\geq$ 20 Myr, 
$1.1\times 10^9\leq M\leq 4.8\times 10^{10}M_\odot$). However, we do not have any
constraint on either $Z$ or $\tau$.

   {\bf Object \#13:} This object has a best fit of $\widetilde\chi^2=1.0$
with parameters of 
($z_p=4.7$, $T=1.3$ Gyr, $\tau=0.4$ Gyr, $M=1.8\times 10^{10} M_\odot$,
$E(B-V)=0$, $Z=1/200Z_\odot$). High-quality solutions are available only when
($Z<Z_\odot$, $E(B-V)\leq 0.23$ mag, $T>0.5$ Gyr, $\tau\geq 0.2$Gyr,
$0.8\times 10^{9}\leq M\leq 3.8\times 10^{10} M_\odot$).

   {\bf Object \#14ab:} The best-fit to this object
has $\widetilde\chi^2=0.11$, and its parameters are
($z_p=4.5$, $T=0.7$ Gyr, $\tau=0.2$ Gyr, $M=2.1\times 10^{10} M_\odot$,
$E(B-V)=0$, $Z=Z_\odot$). High-quality solutions can only be found
when ($E(B-V)\leq 0.23$ mag, $0.09\leq \tau\leq 0.8$Gyr, $T\geq 80$ Myr,
$1.3\times 10^{10}\leq M\leq 7.8\times 10^{10} M_\odot$). However, we do
not have constraint on its metallicity.

   {\bf Object \#20:} This object can also be very well fitted. The best-fit
has $\widetilde\chi^2=0.14$ with parameters of
($z_p=4.9$, $T=0.9$ Gyr, $\tau=0.7$ Gyr, $M=3.7\times 10^9 M_\odot$,
$E(B-V)=0$, $Z=1/200Z_\odot$). However, we do not have any constraint on
either $T$ or $\tau$, as high-quality solutions can be found at any values.
The metallicity is also unconstrained, but there is evidence that the 
goodness-of-fit improves towards lower $Z$. The estimate on stellar mass spans
a wide range, from $5.0\times 10^8 $ to $1.0\times 10^{10} M_\odot$. The
only tightly constrained parameter is the amount of reddening, for which we
find $E(B-V) \leq 0.23$ mag in order to find high-quality solutions.

   {\bf Object \#51abcde:} The best-fit to this rich group has following
parameters: 
($z_p=4.6$, $T=1.0$ Gyr, $\tau=0.4$ Gyr, $M=1.0\times 10^{10} M_\odot$,
$E(B-V)=0$, $Z=1/200Z_\odot$). We do not have constraint on either $\tau$ or
$Z$, but find that the goodness-of-fit improves towards lower $Z$. We can find
high-quality solutions only if ($E(B-V)\leq 0.23$ mag, $T\geq 50$ Myr, 
$2.0\times 10^9 \leq M \leq 2.2\times 10^{10} M_\odot$).

   The analysis above also shows that galaxies with stellar masses of order
$M\sim 10^{10}M_\odot$ existed at $z\approx 5$, and that they likely formed at
much earlier epochs. Similar to the $z\approx 6$ objects, the reddening in
these $z\approx 5$ galaxies is moderate or even negligible. While the fits to
the SEDs of a number of galaxies show a trend of improved quality towards lower 
metallicities (counter-intuitively as compared to the results in the 
$z\approx 6$ sample), they are all consistent with solar abundance.

\subsection{Blue rest-frame UV colors in the $z\approx 5$ sample}

   Similar to the case of the blue rest-frame UV colors in the $z\approx 6$ 
sample discussed in \S 5.1, we find that all the $z\approx 5$ objects
have blue $(i_{775}-z_{850})$ colors indicative of recent star-formation, and
that some of them also have abnormally blue $(i_{775}-z_{850})$ colors that
are hard to explain even the zero-age models of BC03. This is
illustrated in Fig. 6, where the observed data points are superposed on the
``isochrone" tracks like those shown in Fig. 3. The four objects with
confirmed redshifts are the squares without horizontal error bars, while the
remaining objects have redshift error bars indicating a possible range of 
$4.5\leq z\leq 5.5$.

    The objects that have the largest deviations from the models are \#4ab
and \#15ab, which are at $z=5.01$ and $5.49$, respectively. The discrepancies,
however, are different for these two objects, and might represent two different
mechanisms. Object \#4ab deviates from the models similarly to the object
\#1ab in the $z\approx 6$ sample, and such a discrepancy might be due to the
high-mass end cut-off of the IMF adopted by BC03. The blue color of object
\#15ab, on the other hand, cannot be explained in this way. Although it might
be reproduced if the galaxy redshift were significantly smaller, we note (\S 3.4) 
that the redshifts from both Keck and the GRAPES ACS grism spectrum agree very
well ($z = 5.40$ and 5.52, respectively), so we consider this explanation unlikely.
Given its redshift, its intrinsic UV emission in the $i_{775}$-band, no matter how
strong it is, would be easily quenched by the \Lya absorptions of the IGM HI 
clouds along the sight-line. The fact that its observed $i_{775}$-band flux is
much stronger than expected suggests that the IGM HI  absorption along this
particular sight-line might not be as strong as what Madau (1995) prescribed.
If we keep the absorption scheme of Madau (1995), the photometric redshift
for this object would be well constrained at $z_p=4.60$ instead of the
spectroscopic redshift $z=5.49$. A toy model that offers the simplest 
explanation to its observed color, therefore, is that the distribution of the
IGM HI  clouds still follows
the law of Madau (1995) all the way to $z\approx 4.6$, but there are very few
H I clouds from $z\approx 4.6$ to 5.49. We emphasize that such a toy model
might not be physical, and that a plausible model should wait until
sufficiently high S/N, high resolution spectrum is available for this object.

\section{The number density of massive galaxies at $z \approx 5$ to 6}

  The analysis presented here shows that galaxies with stellar masses
$M \gtrsim 10^{10} M_\odot$ existed at $z \approx 5$ to 6, when the
universe was only $\sim 1$~Gyr old.  Moreover, some of those galaxies
appear to have been forming stars for several hundred Myr, beginning
at quite large redshifts, $z \approx 7$ to 20.  Eyles et al.\ (2005)
derived similar properties for the two $z \approx 6$ galaxies that
they analyzed.   The stellar masses of these galaxies are similar to
those that have been estimated for typical (``$L^\ast$'') Lyman break
galaxies at $z \approx 3$ (Papovich et al.\ 2001), when the universe 
was roughly twice as old as it was at $z \approx 5$--6.  As we have 
noted in \S3.1, the high-$z$ galaxies that are detected by IRAC are also
among the optically brightest star-forming galaxies at these redshifts,
and thus may be at the upper end of the mass function, although this 
has yet to be properly quantified.  We may ask, however,
if the existence of objects with these masses is consistent with
predictions from models of galaxy formation in a $\Lambda$--dominated
cold dark matter ($\Lambda$CDM) universe.  There has been a lively
debate in the literature as to whether such models can reproduce the
observed number density and stellar population properties of massive
galaxies at lower redshifts ($z \approx 1$ to 3, e.g.,
Fontana et al.\ 2003; Poli et al.\ 2003; Nagamine et al.\ 2004;
Somerville et al.\ 2004).  While our HUDF+IRAC sample is not complete
in any sense, it is still useful to compare the lower limits on
massive galaxy space density thus derived against the models in order
to extend these tests to earlier cosmic times.

  For this purpose, we compare our data to recent hydrodynamic simulations
in a $\Lambda$CDM universe as described in Nagamine et al.\ (2004)
and Night et al.\ (2005).  Those authors kindly provided us with mass
functions from two types of simulations, namely, a Smoothed Particle
Hydrodynamics (SPH) simulation and a Total Variation Diminishing (TVD)
simulation.  The SPH model simulates a co-moving volume nearly 100 times
larger than that of the TVD model (box sizes of 100$h^{-1}$ Mpc and
22$h^{-1}$ Mpc, respectively), and therefore provides better statistics
for rarer, high mass objects.  In the mass range of overlap, however,
and before resolution effects limit the SPH simulation, the TVD model
predicts roughly two to three times as many galaxies per unit volume
at fixed stellar mass.
 
  Our galaxy sample is by no means complete due to the fact that we
have considered only objects that were reasonably isolated in the IRAC
images and therefore whose detection and photometry was not subject
to blending issues.  We may nevertheless estimate the stellar mass 
limit at which our sample should be complete, modulo IRAC crowding.
We have not rigorously quantified the effects of incompleteness.  
However, in \S 3.1 we have noted that this relatively isolated, 
IRAC-detected subsample makes up about one third of the total number 
of galaxies in each color--selected sample range down to the ACS 
magnitude limits of 26.9 in $z_{850}$ (for $z \approx 6$) or $i_{775}$ 
(for $z \approx 5$), and that only a few of the other two-thirds fraction 
not analyzed here are clearly {\it undetected} in the IRAC data.  
We infer, therefore, that the incompleteness correction in the 
IRAC-detected sample is unlikely to be more than a factor of 
$\sim 3$.  
 
  As described in \S 3.1, object detection in IRAC was done on the
weighted sum of the 3.6$\mu$m and 4.5$\mu$m images.  The 3.6$\mu$m
data are deeper and contribute most of to the signal-to-noise ratio
of this sum, and therefore we consider that band only for the IRAC
detection limit.  While the formal $S/N = 5$ point source limit of
the GOODS IRAC data is $m_{3.6\mu m} = 26.4$ mag, we will adopt a brighter
$S/N = 10$ limit $m_{3.6\mu m} = 25.65$ mag for the present analysis,
where detection and photometry are more secure, and where the effects
of crowding and incompleteness are reduced.  All but one of the galaxies 
in the subsample for which we have derived mass estimates (\#20 at 
$z_p = 4.9$) are brighter than this limit.  For the redshift range covered 
by the ACS color selection (roughly $4.5 \leq z \leq 6.5$), the 3.6$\mu$m 
band measures rest-frame light in the 0.48-0.65$\mu$m rest frame
wavelength range, i.e., on average around the rest-frame $V$--band.
Galaxies in the sample must have sufficient ongoing star formation
to produce the UV light which allows them to be detected by ACS.
Their ages, star formation histories, and dust content result in
different stellar mass-to-light ratios ($M/L$).  In practice, for
the stellar population models which best fit the objects in our
sample, we find $0.17 < M_\ast/L_V < 0.66$ in solar units.  The
upper bound is not far from the maximum $M/L_V$ that we would
expect for an unreddened stellar population at this redshift,
given the age of the universe (1 Gyr at $z = 5.8$, yielding
$(M/L_V)_{max} = 0.77$).  Therefore, we adopt the upper bound 
of the model-fit $M/L$ range as appropriate for the sort of
UV-selected objects in this sample, but it would make little
difference to use the maximum allowable $M/L$.  At $z = 6.5$,
for this upper bound to $M/L$, our adopted IRAC magnitude limit
corresponds to a stellar mass limit of $1.6 \times 10^{10} M_\odot$.
We therefore expect to be able to detect any galaxy above this mass
limit in the IRAC data at these redshifts, and we adopt this
threshold for comparison to the models.
 
To calculate the effective volumes over which we identify galaxies
in our sample, we use the solid angle of the HUDF, and the redshift
selection function for our color criteria (\S 2), which we evaluate
from Monte Carlo simulations following the procedures described in
Giavalisco et al. (2004b).  We selected galaxies within the clean area
of the ACS HUDF image that remains after trimming off the noisy edges,
and which subtends 10.34 arcmin$^2$.    In the Monte Carlo simulations,
artificial galaxies with realistic distributions of UV rest-frame
colors, sizes, and magnitudes, and spanning the range of redshifts
of interest, were added to the real HUDF ACS images and recovered
with SExtractor.  The fraction of input objects that are recovered
and which meet the color selection criteria, as a function of redshift,
gives the redshift selection function, which we then integrate to get
the effective volume for the sample.  Within the HUDF solid angle,
our selection criteria correspond to effective survey volumes of
$2.5\times 10^4$ Mpc$^3$ and $2.8\times 10^4$ Mpc$^3$ for the
$z \approx 6$ and $z \approx 5$ samples, respectively.
 
We use the best-fit stellar masses for the galaxies we have analyzed, 
as reported in Tables 3 and 4.  In the $z\approx 6$ sample, there are 
two galaxies (out of three total) with $M>1.6\times 10^{10} M_\odot$, 
corresponding to a space density of $0.8\times 10^{-4}$ Mpc$^{-3}$ for 
this mass range.  In the $z\approx 5$ sample, there are four galaxies 
(out of eight with derived masses) with $M>1.6\times 10^{10} M_\odot$, 
or a space density of $1.4\times 10^{-4}$ Mpc$^{-3}$.  For comparison, 
at $z = 6$, the SPH and TVD simulations give cumulative number densities 
of $2.1\times 10^{-4}$ Mpc$^{-3}$ and $3.7\times 10^{-4}$ Mpc$^{-3}$,
respectively, or 2.6 and 4.6 times larger than the observed
number.  At $z = 5$ the number densities in the models are
$9.0\times 10^{-4}$ Mpc$^{-3}$ and $18.6\times 10^{-4}$ Mpc$^{-3}$,
or 6.4 and 13.3 times larger than the number of galaxies in our
sample above this mass threshold.
 
Even allowing for a generous degree of incompleteness in our sample,
we conclude that $\Lambda$CDM models like those considered here can
successfully produce enough galaxies at these redshifts with masses like 
those we have inferred.  The basic nature of the result is largely
insensitive to the choices we have adopted such as the IRAC magnitude
limit or the maximum allowable $M/L$ threshold, unless there were a 
substantial number of comparably massive galaxies with UV fluxes so 
faint as to be undetected in the ACS HUDF images, or, at least, not 
selected by our color criteria.  The actual ratios of the observed 
to predicted number densities are, however, quite sensitive to these 
details, due to the very steep mass function in the simulations 
($dN/dM_\ast \propto M_\ast^\alpha$, with $\alpha \approx -2.5$ in 
this mass range, which we note is divergent if extended to arbitrarily 
low masses, below the resolution of the simulations).  We also note 
that these models predict quite steep redshift evolution in the number
density of objects in this mass range (roughly a factor of 4 to 5 
increase from $z = 6$ to $z = 5$).  Future analysis of the GOODS 
{\it Spitzer} data should allow us to test these predictions by examining 
complete, infrared--selected samples that will more nearly approximate
populations limited by stellar mass at different redshifts.
Two ultradeep IRAC fields have also been observed in GOODS-N,
which will provide additional dynamic range to help test the
steepness of the mass function predicted by the simulations.
Analysis of other data will also help to control the potential
field-to-field variations in the observed number density that
may result from clustering, which we may expect would be
strong for galaxies at the upper range of the mass function
at these redshifts.

\section{Summary}

   Using the first epoch of GOODS IRAC observations, we have identified 
mid-infrared counterparts of the $z\approx 6$ and 5 galaxy candidates selected 
in the HUDF.  Six of these galaxies have spectroscopic confirmation, verifying 
that the IRAC instrument can indeed probe galaxies to such high redshifts, 
and as faint as $\sim 26.9$ mag in the ACS $z_{850}$ or $i_{750}$ bands.  
In this paper, we study three and eleven objects in the $z\approx 6$ and 5 
samples, respectively, all of which are reasonably isolated so that crowding 
in the IRAC images does not compromise their photometry.

   Combining the photometry from ACS, NICMOS (in one occasion also ISAAC 
at the VLT) and IRAC, we analyze the rest-frame UV to optical spectral energy
distributions of these galaxies and compare them to stellar population synthesis 
models.  IRAC samples redshifted rest-frame optical wavelengths where there 
is a greater contribution of light from intermediate-age and long-lived 
stars, and where the effects of dust extinction are reduced compared to those 
in the ultraviolet rest frame.  These data therefore help us to estimate the 
stellar masses and ages of these galaxies.  Deep near-infrared data in two 
passbands from NICMOS are also available for many of the objects.  These 
provide measurements of the UV luminosity and spectral slope, which are 
sensitive to the on-going star formation rate and degree of dust reddening,
quantities that are insufficiently constrained from the ACS optical data 
alone.  

    While the SED analysis reveals diversity in the stellar populations
of these galaxies, it also demonstrates a number of common properties. 
The following are the most important results from this study.

   (1) Galaxies as massive as $\sim 10^{10} M_\odot$ already existed when
the universe was about a billion years old.  These stellar masses are similar 
to those of typical ($L^\ast$) Lyman-break galaxies at $z \approx 3$, when the 
Universe was roughly twice as old as it was at $z \approx 5$ to 6.  Two out 
of the three objects in the $z\approx 6$ sample, and at least four out eleven 
objects from the $z \approx 5$ sample, have best-fit stellar masses 
$M>1.6\times 10^{10} M_\odot$.   While the acceptable fits to these objects 
span a considerable range in the stellar population model parameter space, 
the lower bound to their stellar masses it well constrained to be 
$> 10^{10} M_\odot$.  Although the galaxy samples are incomplete, the resulting 
lower limits on their space density at these stellar masses can be comfortably 
accommodated by at least one set of recent $\Lambda$CDM models for galaxy 
formation (Nagamine et al. 2004; Night et al. 2005).  Those models predict 
a very steep mass function and quite rapid redshift evolution of the number 
density of galaxies at fixed stellar mass, predictions which may be 
tested by the analysis of larger and deeper samples from the GOODS 
{\it Spitzer} data set.

   (2) The photometry for most of the galaxies studied here shows evidence for
a pronounced increase in flux density between the rest-frame UV wavelengths
(ACS and NICMOS) and the optical light measured by IRAC.  This is naturally 
reproduced by a Balmer break that results from the dominant presence of stars 
with ages of a few hundred Myr.  All three $z\approx 6$ objects have best-fit 
ages $\gtrsim 0.1$ Gyr, while at least eight $z\approx 5$ objects have best-fit 
ages $\gtrsim 0.4$ Gyr. In particular, the $z\approx 6$ sample strongly indicates 
that the universe was already forming galaxies as massive as $\sim 10^{10} M_\odot$ 
at $z\gtrsim 7$, and possibly even at $z\sim 20$.  First-year results from WMAP 
suggest that the reionization of the universe first began at $z\sim$ 15--20
(Spergel et al.\ 2003).  The stellar population properties that we derive are 
in qualitative agreement with this picture, demonstrating that stars, which 
provide an important -- if not the only -- source of reionizing photons, 
could indeed be formed at such an early epoch.

   (3) While there is no firm constraint on the metallicity of these galaxies, 
all the objects analyzed here are consistent with solar abundance, which 
indicates that the galaxies in such an early stage of the universe might have 
already been significantly polluted by metals.  In fact, the best-fit models to 
all the three $z\approx 6$ objects have solar metallicity, and for one object 
there is strong evidence that its metallicity must be higher than $1/200 Z_\odot$. 
If these metallicities are representative at $z \approx 6$, this may cast doubt 
on the suggestion ({\it e.g.}, Stiavelli et al. 2004) that extremely metal-poor 
galaxies could be the major sources that finished the reionization at $z\approx 6$. 

   (4) In all cases, the best-fit stellar population models have no dust 
reddening, and the allowed models generally have low extinction values.  
Even so, several galaxies, including three with spectroscopic confirmation, 
show abnormally blue rest-frame UV colors compared to these unreddened models.
We find that in two cases (at $z=5.83$ and $z=5.05$), an IMF extending to 
very massive stars ($>100M_\odot$), which are not included in the models
used here, might account for such blue colors.  This would have important 
implications for reionization at $z\approx 6$, offering further evidence 
to justify a large Lyman continuum photon escape fraction that has been
adopted in a number of studies ({\it e.g.}, Yan \& Windhorst 2004a; 
Yan \& Windhorst 2004b).  In another case (at $z=5.49$), the blue color 
might be explained by a reduced degree of intergalactic HI absorption along 
the line of sight compared to the mean opacity predicted by Madau (1995).  

\acknowledgments

   The authors thanks Dr. Nagamine for providing his most recent numerical
simulations. We thank the other members of the GOODS team who have contributed
to the success of the observations and data analysis. We also thank the referee 
for very helpful comments. Support for this work, part of the 
{\it Spitzer Space Telescope}
Legacy Science Program, was provided by NASA through Contract Number 1224666
issued by the Jet Propulsion Laboratory, California Institute of Technology
under NASA contract 1407.

\begin{figure}
\plotone{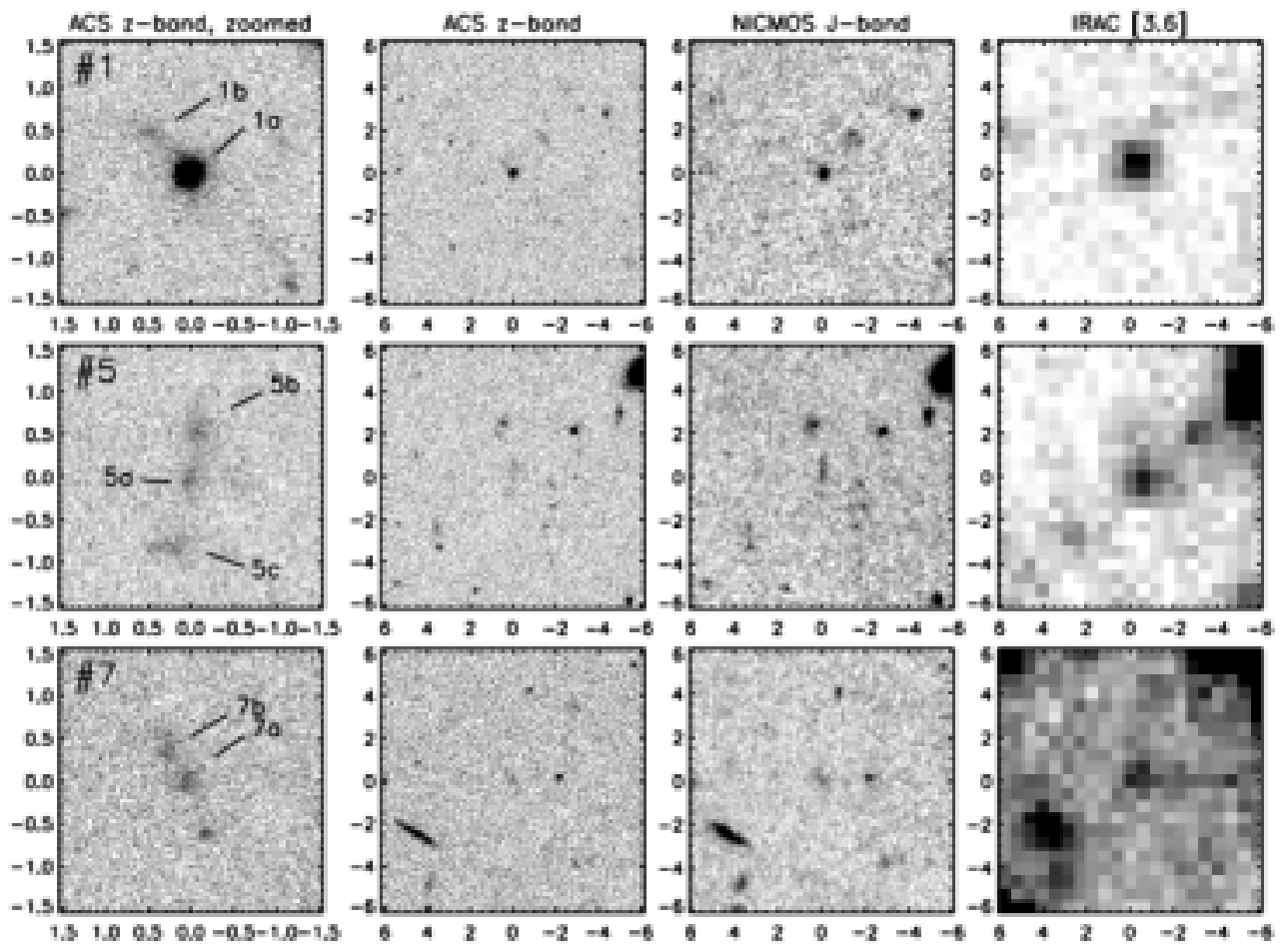}
\caption{Image stamps of the IRAC-detected $z\approx 6$ objects.
The ACS $z_{850}$-band ``stamps" in the left panel are $1.5''\times 1.5''$ in
size, and show the morphological details of these objects. The images in the
other three panels ($12''\times 12''$ in size) shows the matching of these
sources from ACS $z_{850}$ to IRAC $3.6\mu m$. NICMOS $J_{110}$ stamps are
also shown when available. The ID's of these objects are taken from YW04.
Object \#1ab has a spectroscopically measured redshift $z = 5.83$.
}
\end{figure}

\clearpage
\begin{figure}
\setcounter{figure}{1}
\plotone{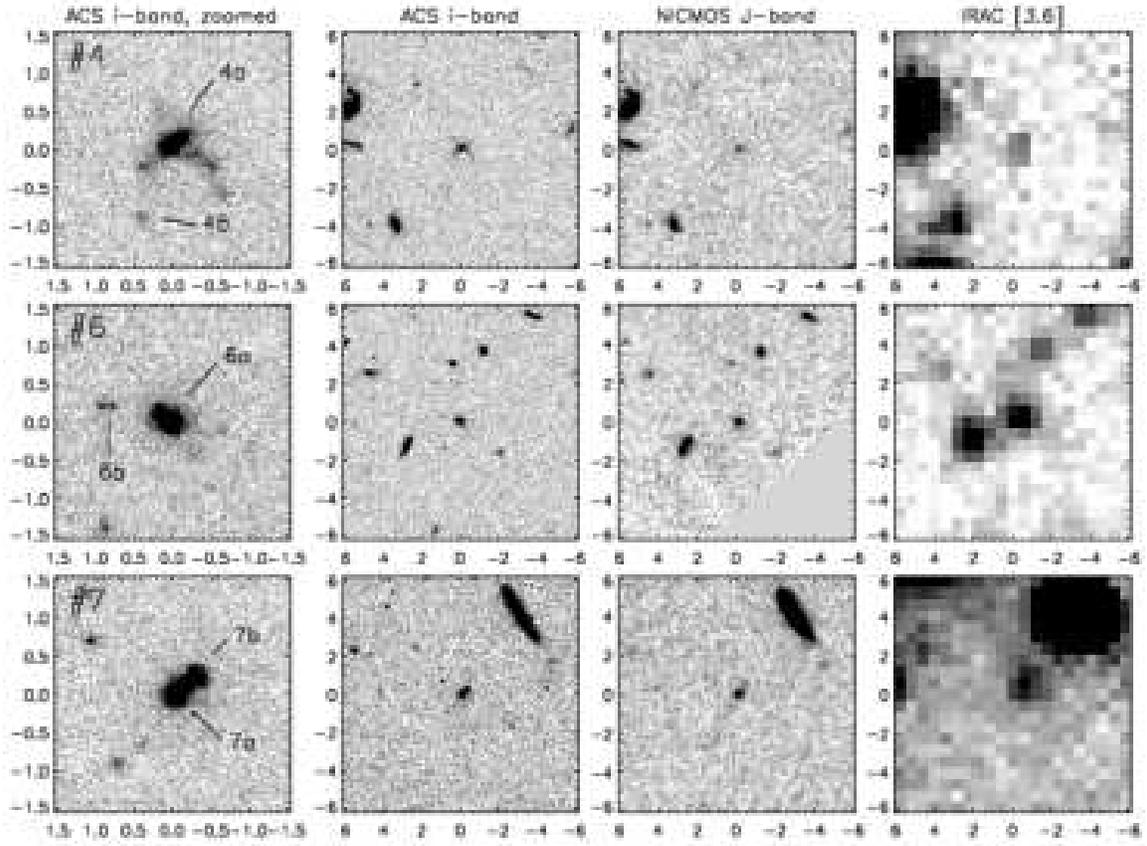}
\caption{Image stamps of the IRAC-detected $z\approx 5$ objects.The ID's
are taken from Yan et al. (in preparation).
}
\end{figure}

\setcounter{figure}{1}
\begin{figure}
\plotone{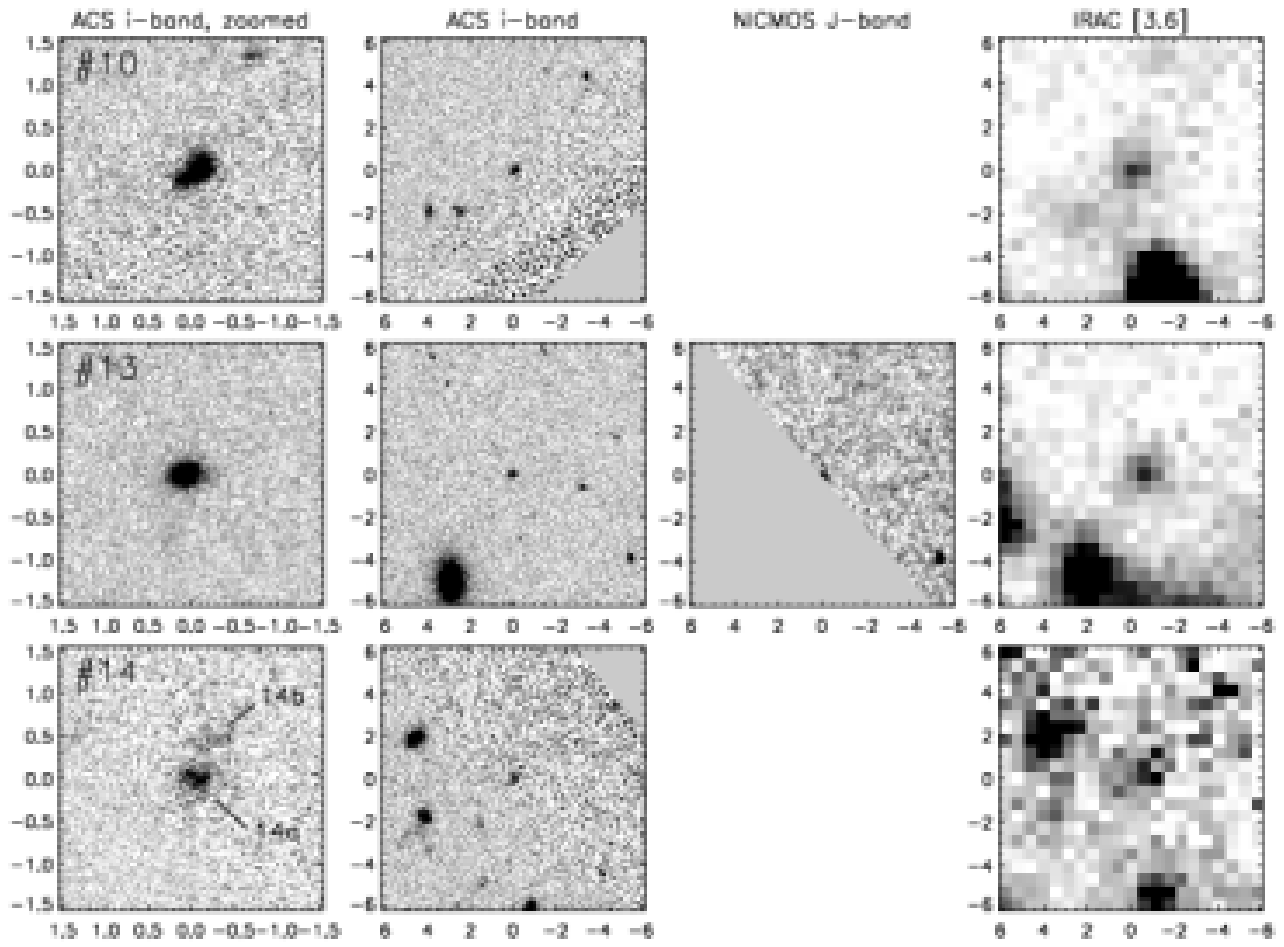}
\caption{Continued.
}
\end{figure}

\setcounter{figure}{1}
\begin{figure}
\plotone{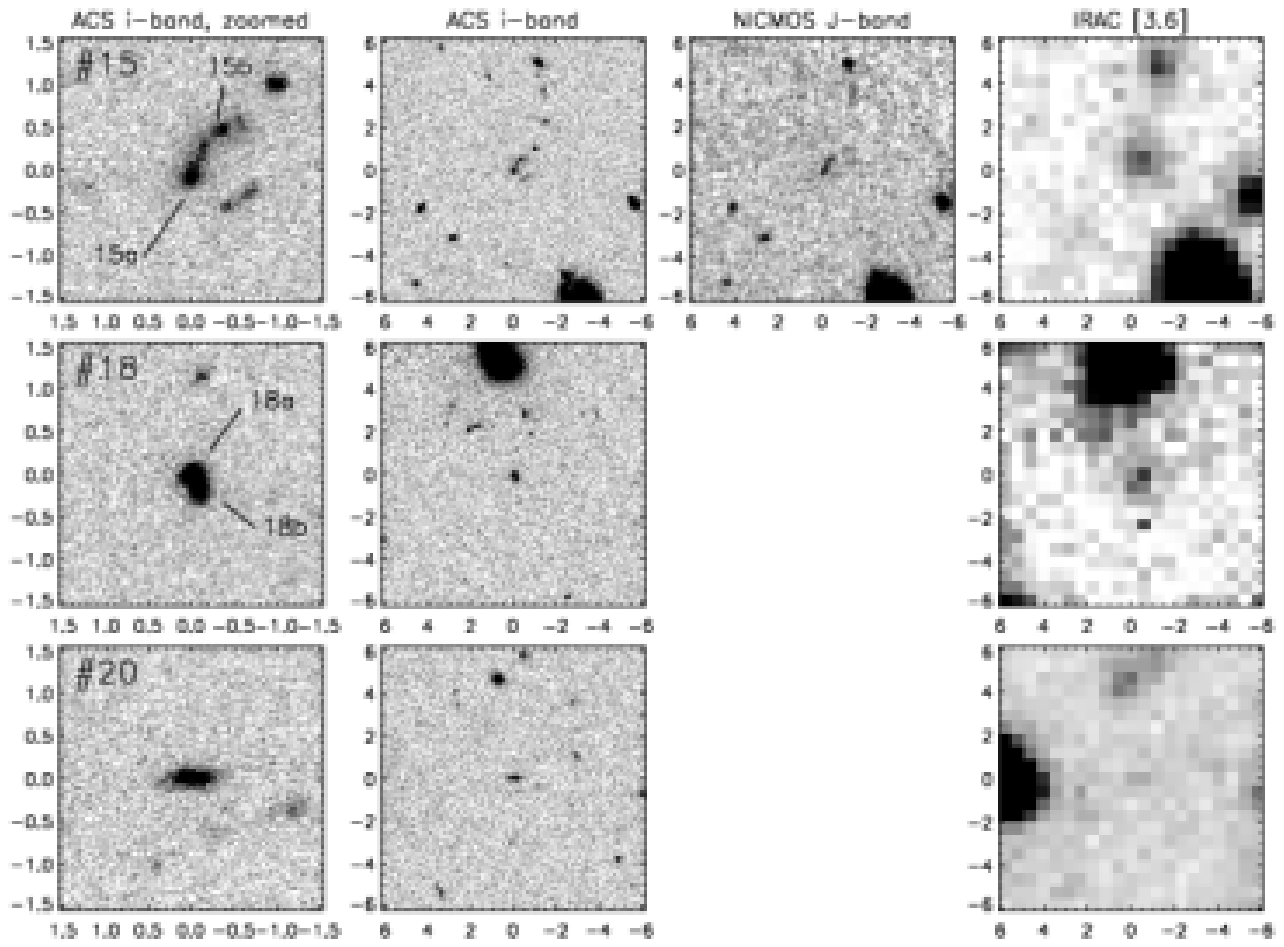}
\caption{Continued.
}
\end{figure}

\setcounter{figure}{1}
\begin{figure}
\epsscale{1.0}
\plotone{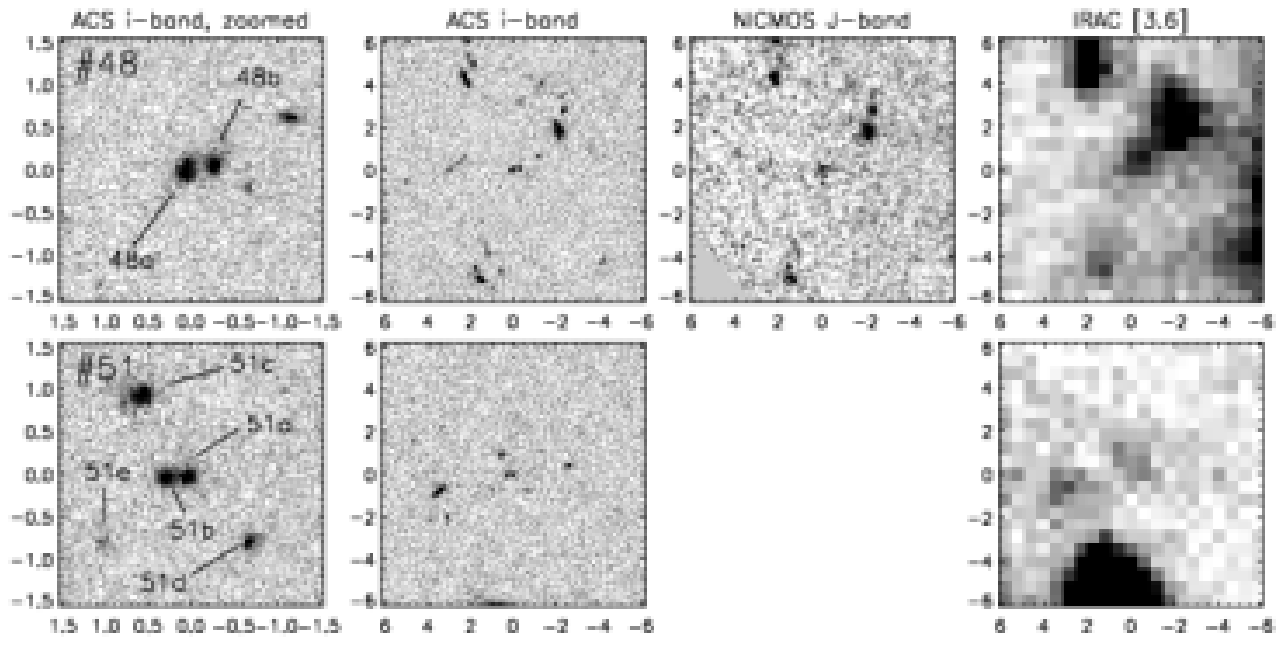}
\caption{Continued.}
\end{figure}

\clearpage

\begin{figure}
\setcounter{figure}{2}
\plotone{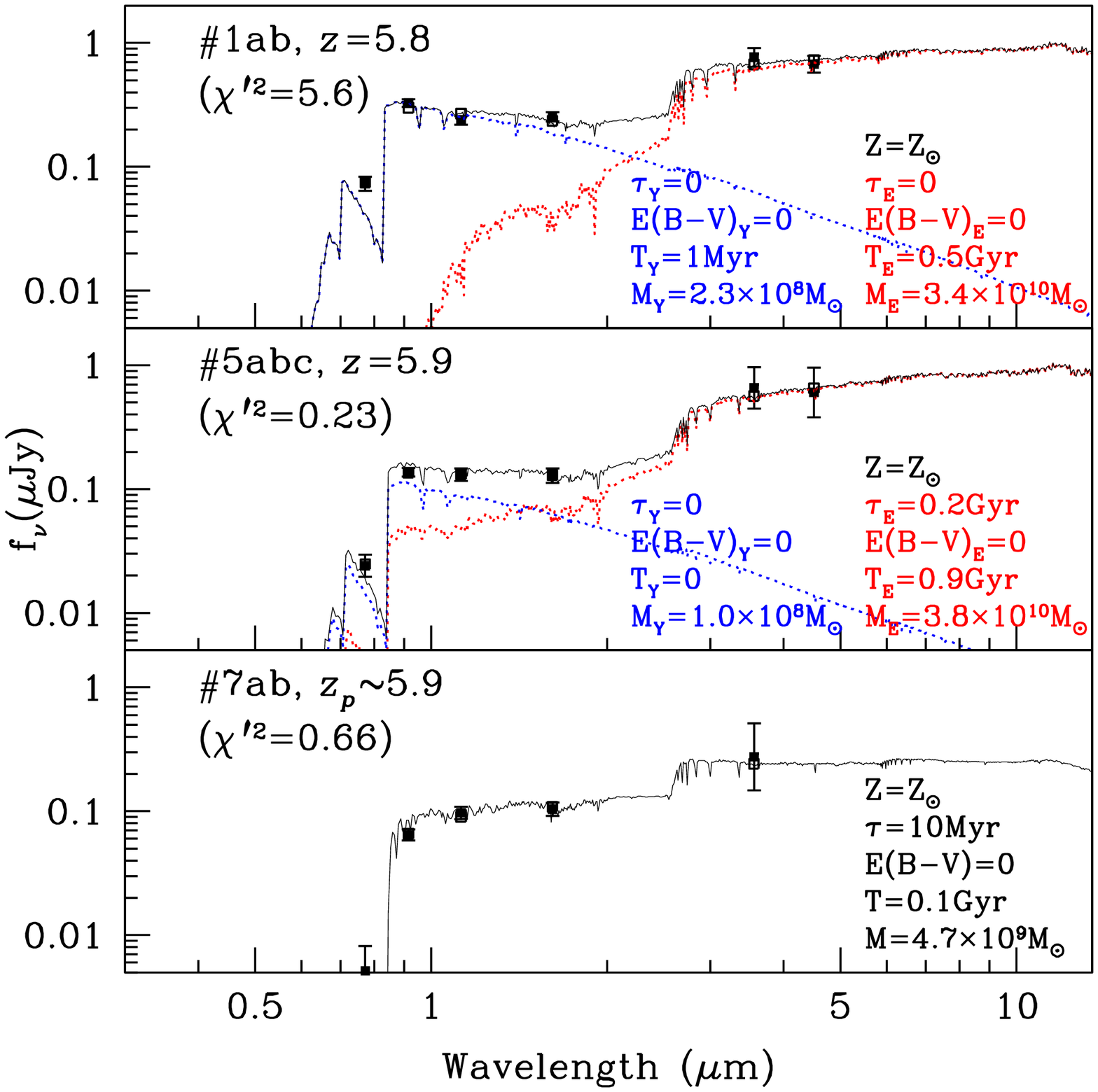}
\caption{The spectra of the best-fit models to the three IRAC-detected
$z\approx 6$ objects are shown in this figure. The parameters of these models
are also labeled. The observed SEDs are plotted as filled squares with error
bars. The model spectra are overplotted, and the synthetic model photometry
integrated through the bandpasses is shown as open squares which often overlap
with the filled squares.
While \#7ab can be explained by 
single-component models, both \#1ab and 5abc require two-component models.
For these two objects, the templates of the evolved and the young components
are shown in red and blue dashed lines, respectively, and their corresponding
parameters are also labeled in red and blue, respectively.
}
\end{figure}

\clearpage
\begin{figure}
\epsscale{.80}
\plotone{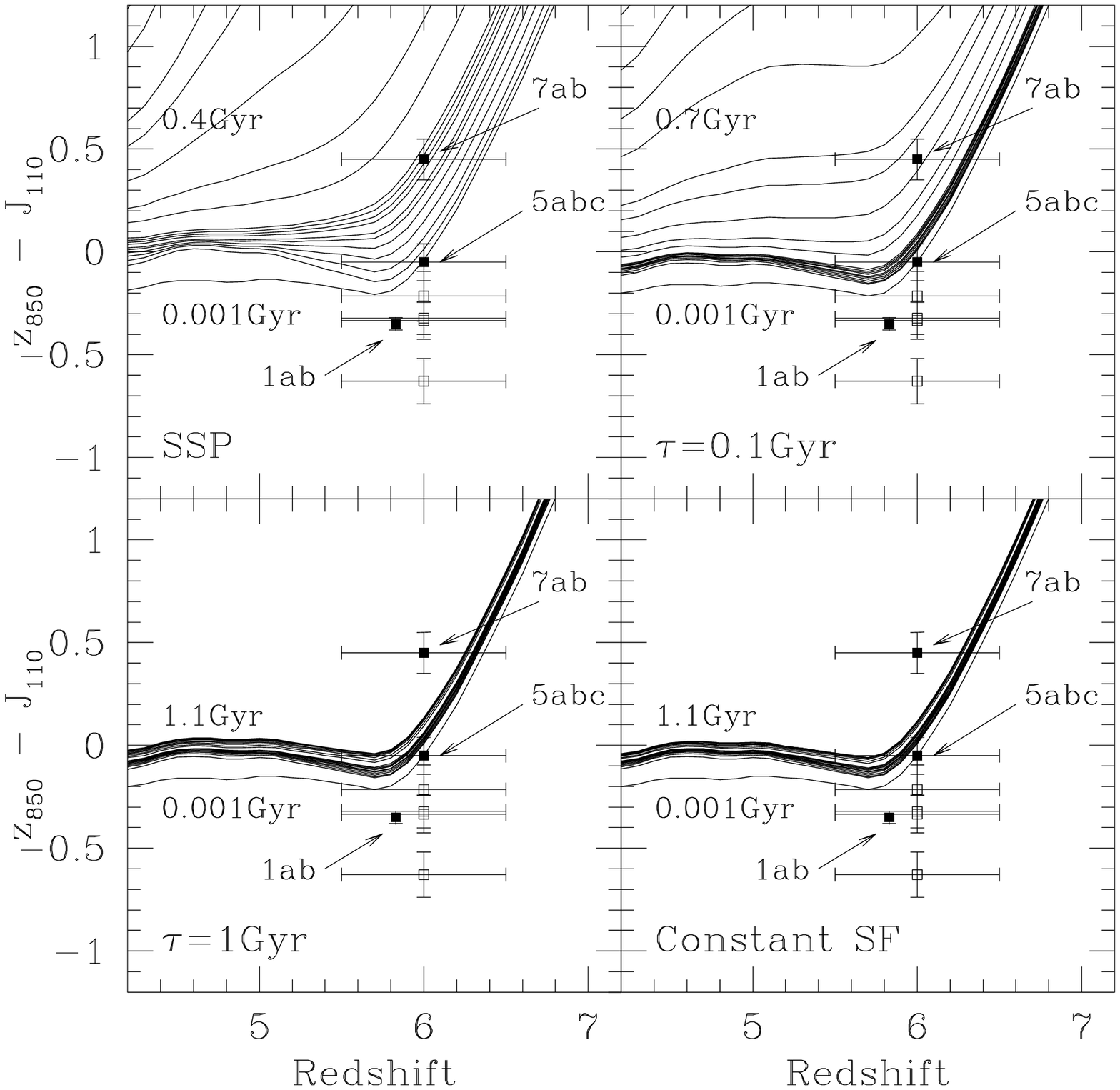}
\caption{Most of the HUDF $z\approx 6$ objects that have NICMOS photometry
show unusually blue $(z_{850}-J_{110})$ colors (YW04). The three that have IRAC
identifications are shown here as filled squares, while the remaining ones are
shown as open squares. The one without horizontal error bar is object \#1ab,
which has a known redshift of 5.83. The others (including \#5abc; see 
explanation in \S 3.4) are put at $z=6$ with error bars indicating their
possible redshift range of $5.5\lesssim z \lesssim 6.5$. The BC03 models are
used to calculate expected colors at different redshifts and ages. Four types
SFH are considered and are shown separately as the 
``isochrones" in each of the four panels: an instantaneous burst (SSP),
continuous but declining star formation with e-folding time scales of 
$\tau=0.1$ and 1 Gyr, and a constant star formation. The ages of the
models are 0.001 Gyr, 0.01 to 0.09 Gyr (0.01 Gyr step-size), and 0.1 to 1.1 Gyr
(0.1 Gyr step-size). 
}
\end{figure}
\clearpage

\begin{figure}
\epsscale{1.0}
\plotone{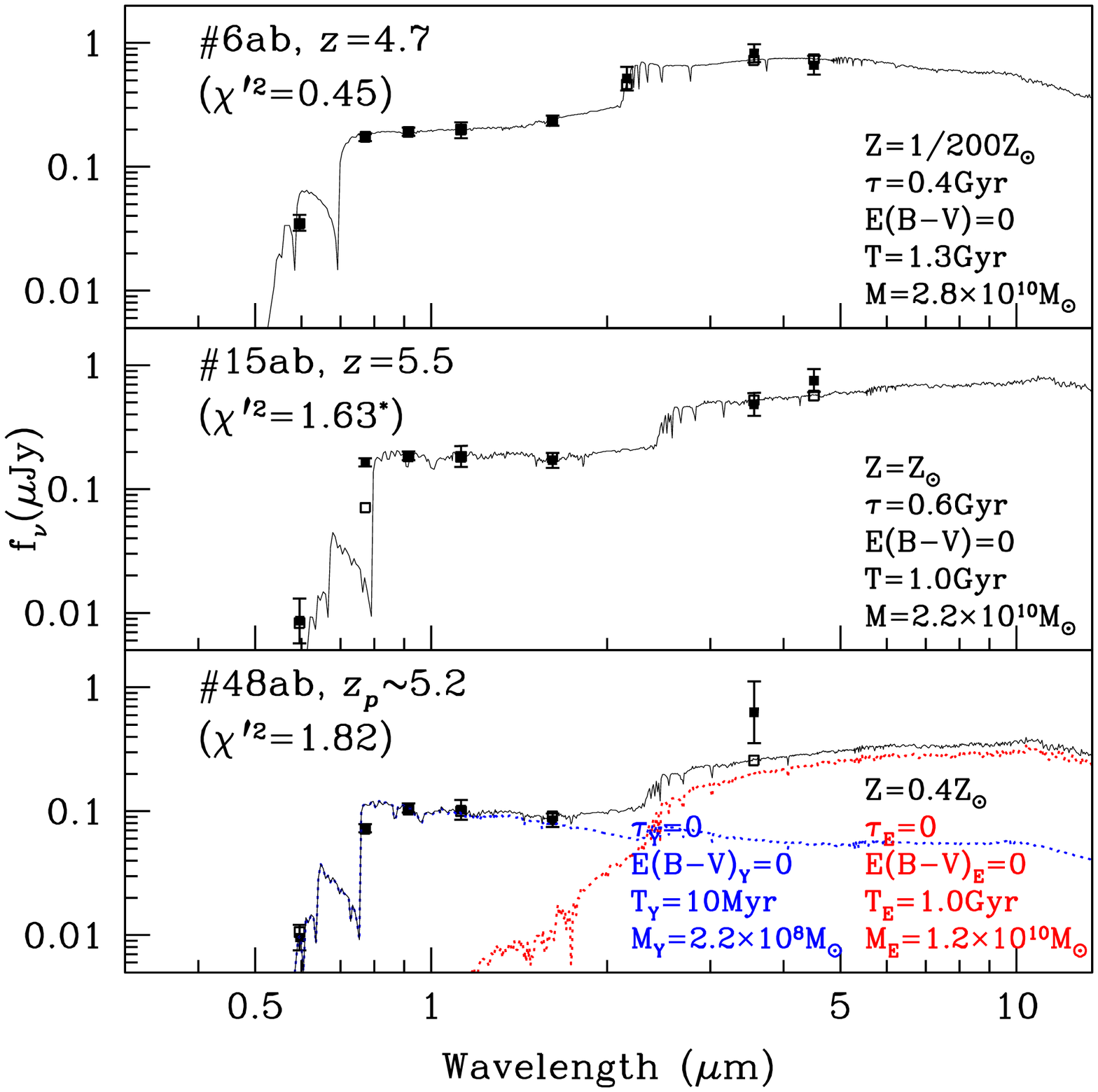}
\caption{The best-fit models to the three $z\approx 5$ objects that have 
NICMOS photometry. Symbols are the same as in Fig. 3. Object \#6ab can be well
fitted by single-component models. Object \#15ab, is problematic because of
its abnormally blue $(i_{775}-z_{850})$ color. If we ignore $i_{775}$-band as
justified in \S 6.2 (see also Fig. 6), reasonably high quality fits can be
obtained using single-component models. Object \#48ab, on the other hand,
seems need two-component models. However, we consider the fit to this object
uncertain (see \S 6.1 for details). 
}
\end{figure}

\begin{figure}
\plotone{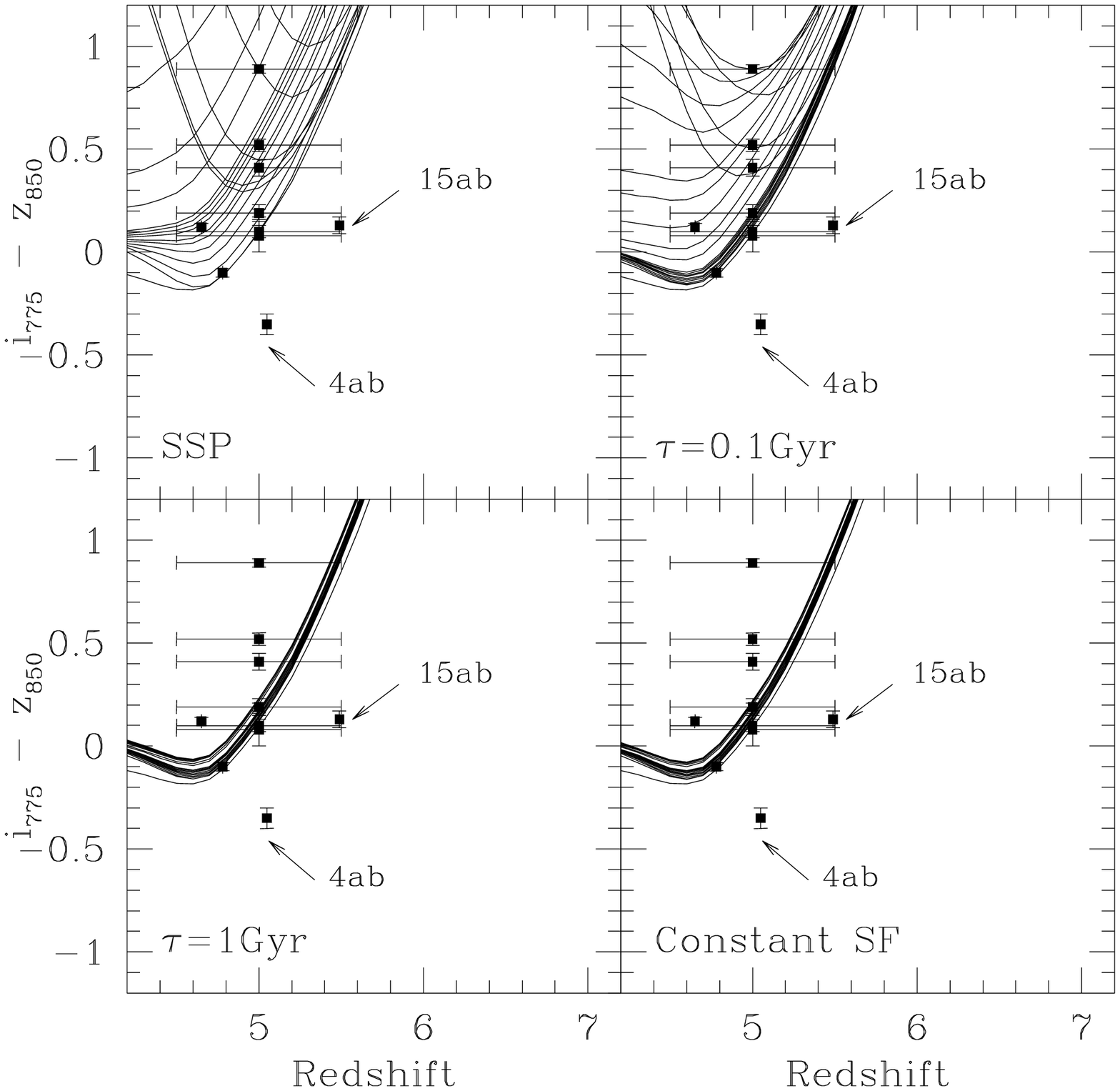}
\caption{Similar to the abnormally blue rest-frame UV colors observed in some
of the $z\approx 6$ candidates, a significant number of $z\approx 5$ objects in
the HUDF also have blue UV colors that are inconsistent with any model. For
clarity, this figure shows {\it only} the $z\approx 5$ objects in our sample.
The ``isochrone" tracks, as well as the symbols, are the same as in Fig. 4. 
The four objects without horizontal error bars are those that have been
spectroscopically confirmed. All other objects are plotted at $z=5$ with error
bars indicating the possible redshift range of $4.5\leq z\leq 5.5$. 
}
\end{figure}

\begin{figure}
\plotone{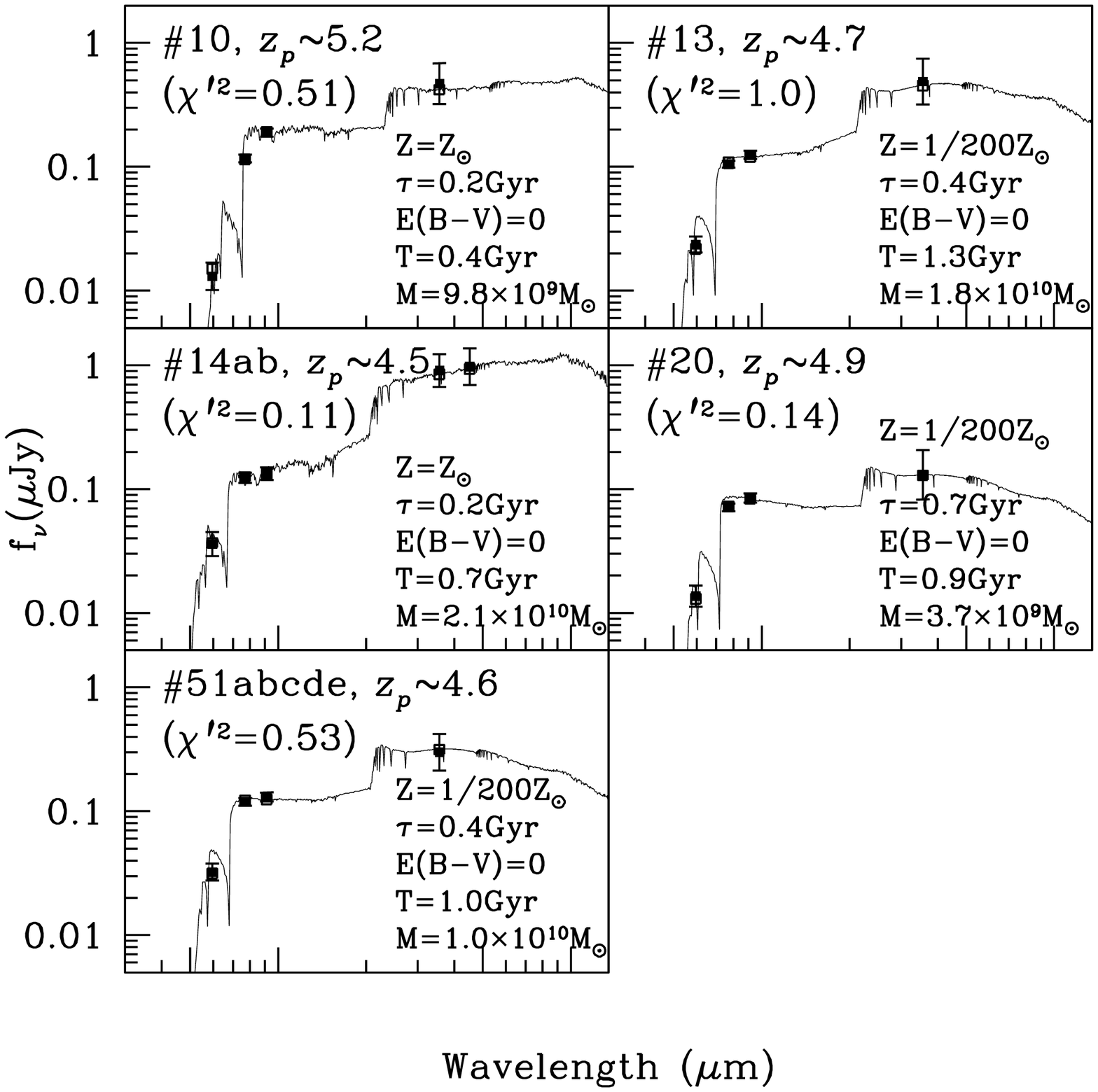}
\caption{The observed SEDs of the $z\approx 5$ objects that have no NICMOS
photometry are shown in this figure, with their best-fit models overplotted.
The symbols are the same as in Fig. 5.
}
\end{figure}

\begin{figure}
\plotone{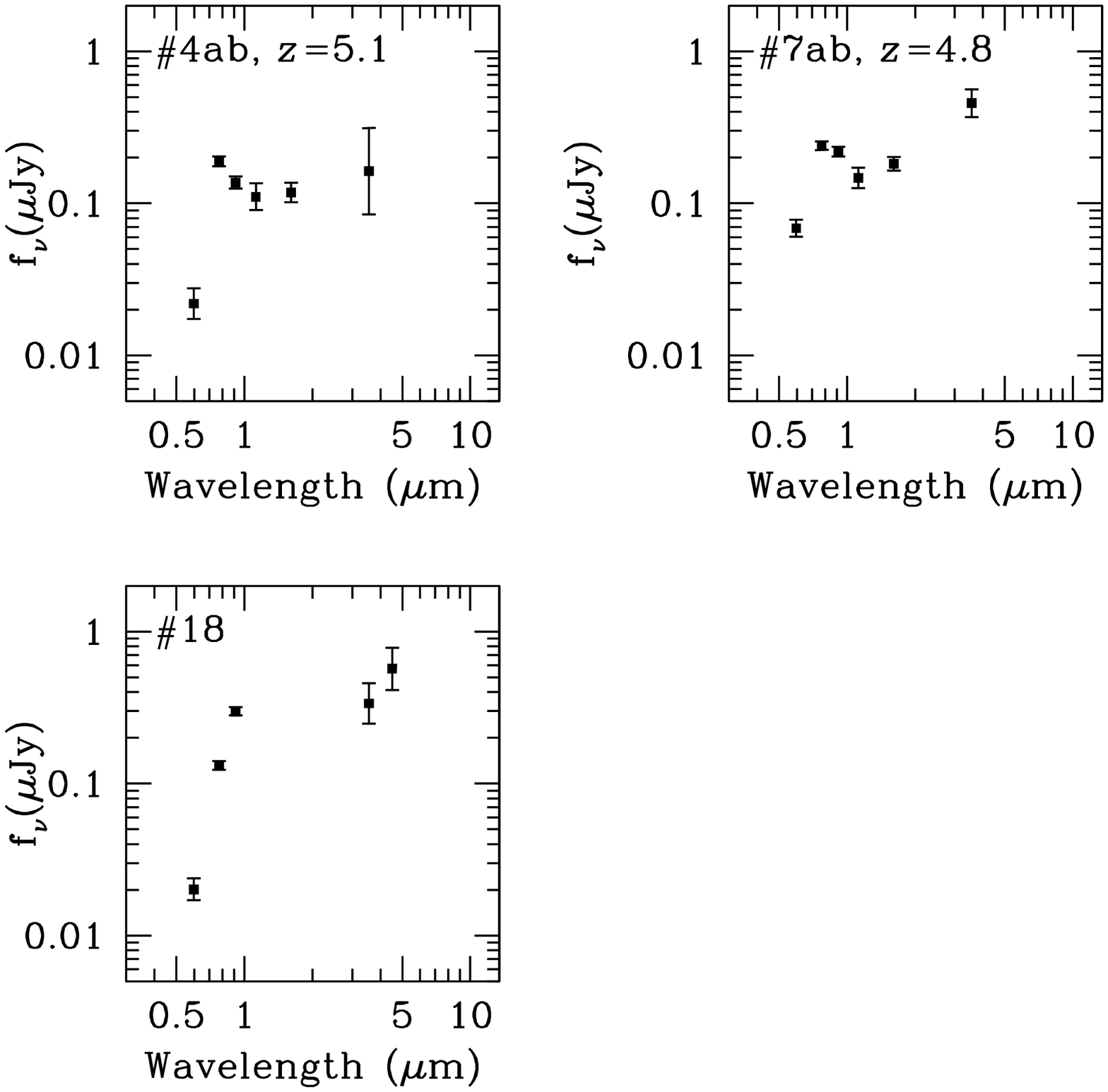}
\caption{Object \#4ab and 7ab in the $z\approx 5 $ sample do not have good
model fit. For completeness, their SEDs are shown here. Also shown is \#18,
which can be fitted by a large range of models and thus is unconstrained.
}
\end{figure}

\clearpage

\begin{deluxetable}{cccccccc}
\tablewidth{0pc}
\tabletypesize{\scriptsize}
\tablecaption{Photometric properties of the IRAC-detected $z\approx 6$ objects\tablenotemark{1}}
\tablehead{
\colhead{ID} &
\colhead{RA \& DEC(J2000)} &
\colhead{$i_{775}$} &
\colhead{$z_{850}$} &
\colhead{$J_{110}$} &
\colhead{$H_{160}$} &
\colhead{$m_{3.6\mu m}$} &
\colhead{$m_{4.5\mu m}$}
}
\startdata

1ab &    ...     &      26.74$\pm$0.14  &      25.11$\pm$0.08  & 25.46$\pm$0.09  & 25.39$\pm$0.10 & 24.19$\pm$0.19 & 24.33$\pm$0.17 \\
1a & 3:32:40.01 -27:48:15.01 & 26.88$\pm$0.03 & 25.25$\pm$0.01  &                &                &                &                \\
1b & 3:32:40.04 -27:48:14.54 & 29.03$\pm$0.18 & 27.41$\pm$0.07  &                &                &                &                \\
5abc &    ...    &      27.95$\pm$0.23  &      26.06$\pm$0.09  & 26.11$\pm$0.13   & 26.13$\pm$0.15 & 24.36$\pm$0.42 & 24.45$\pm$0.50 \\
5a & 3:32:34.29 -27:47:52.80 & 29.07$\pm$0.21 & 26.97$\pm$0.05  &                &                &                &                \\
5b & 3:32:34.28 -27:47:52.26 & 28.98$\pm$0.19 & 27.17$\pm$0.06  &                &                &                &                \\
5c & 3:32:34.31 -27:47:53.56 & 29.42$\pm$0.22 & 27.76$\pm$0.08  &                &                &                &                \\
7ab &      ...   &      29.62$\pm$0.50  &      26.88$\pm$0.11  & 26.43$\pm$0.13  & 26.36$\pm$0.14 & 25.30$\pm$0.68 &          ...   \\
7a & 3:32:37.46 -27:46:32.81 & $> 31.0$ & 27.50$\pm$0.07  &                &                &                &                \\
7b & 3:32:37.48 -27:46:32.45 & 29.70$\pm$0.30 & 27.78$\pm$0.09  &                &                &                &                \\
\hline

\enddata
\tablenotetext{1.}{The combined magnitudes and photometric errors 
are given for these objects. These errors include the systematic
errors (see \S 3.3). For the ACS passbands, the magnitudes
and errors of individual components are also given for reference.
These errors of individual components in the ACS bands are based on
$S/N$ only.}
\end{deluxetable}

\begin{deluxetable}{ccccccccc}
\tablewidth{0pc}
\tabletypesize{\scriptsize}
\tablecaption{Photometric properties of the IRAC-detected $z\approx 5$ objects\tablenotemark{1}}
\tablehead{
\colhead{ID} &
\colhead{RA \& DEC(J2000)} &
\colhead{$V_{606}$} &
\colhead{$i_{775}$} &
\colhead{$z_{850}$} &
\colhead{$J_{110}$} &
\colhead{$H_{160}$} &
\colhead{$m_{3.6\mu m}$} &
\colhead{$m_{4.5\mu m}$}
}
\startdata

 4ab &     ...     &          28.05$\pm$0.25  &      25.71$\pm$0.08  &      26.06$\pm$0.10  & 26.29$\pm$0.22 & 26.22$\pm$0.16 & 25.87$\pm$0.71 &  ...          \\
 4a  & 3:32:41.09 -27:46:42.46 & 28.14$\pm$0.14 & 25.75$\pm$0.02 & 26.12$\pm$0.04 &                &                &                &               \\
 4b  & 3:32:41.12 -27:46:43.32 & 30.80$\pm$0.39 & 29.40$\pm$0.12 & 29.25$\pm$0.18 &                &                &                &               \\
 6ab\tablenotemark{2} &     ...     &          27.53$\pm$0.16  &      25.81$\pm$0.07  &      25.69$\pm$0.08  & 25.66$\pm$0.16 & 25.47$\pm$0.10 & 24.11$\pm$0.18 & 24.35$\pm$0.19\\
 6a  & 3:32:33.98 -27:48:02.05 & 27.63$\pm$0.05 & 25.88$\pm$0.01 & 25.73$\pm$0.02 &                &                &                &               \\
 6b  & 3:32:34.04 -27:48:01.84 & 30.14$\pm$0.29 & 28.87$\pm$0.10 & 29.38$\pm$0.28 &                &                &                &               \\
 7ab &     ...     &          26.81$\pm$0.14  &      25.45$\pm$0.07  &      25.55$\pm$0.08  & 25.98$\pm$0.17 & 25.75$\pm$0.11 & 24.75$\pm$0.23 &  ...          \\
 7a  & 3:32:37.96 -27:47:11.04 & 27.31$\pm$0.04 & 25.90$\pm$0.01 & 26.03$\pm$0.02 &                &                &                &               \\
 7b  & 3:32:37.94 -27:47:10.82 & 27.88$\pm$0.07 & 26.63$\pm$0.02 & 26.66$\pm$0.04 &                &                &                &               \\
 10  & 3:32:31.38 -27:48:13.79 & 28.61$\pm$0.28 & 26.23$\pm$0.08 & 25.71$\pm$0.08 &       ...      &       ...      & 24.72$\pm$0.41 &  ...          \\
 13  & 3:32:41.34 -27:48:43.09 & 27.97$\pm$0.16 & 26.35$\pm$0.07 & 26.16$\pm$0.08 &       ...      &       ...      & 24.68$\pm$0.46 &  ...          \\
14ab &     ...     &          27.51$\pm$0.24  &      26.17$\pm$0.10  &      26.09$\pm$0.13  &       ...      &       ...      & 24.01$\pm$0.33 & 23.93$\pm$0.37\\
14a  & 3:32:37.55 -27:45:20.59 & 27.83$\pm$0.14 & 26.46$\pm$0.05 & 26.38$\pm$0.07 &                &                &                &               \\
14b  & 3:32:37.52 -27:45:20.12 & 29.01$\pm$0.26 & 27.77$\pm$0.09 & 27.67$\pm$0.14 &                &                &                &               \\
15ab &     ...     &          29.06$\pm$0.45  &      25.86$\pm$0.08  &      25.73$\pm$0.09  & 25.74$\pm$0.21 & 25.82$\pm$0.15 & 24.69$\pm$0.23 & 24.21$\pm$0.23\\
15a  & 3:32:33.27 -27:47:24.94 & 29.45$\pm$0.34 & 26.50$\pm$0.03 & 26.26$\pm$0.04 &                &                &                &               \\
15b  & 3:32:33.24 -27:47:24.50 & 30.37$\pm$0.84 & 26.75$\pm$0.03 & 26.76$\pm$0.06 &                &                &                &               \\
18ab &     ...     &          28.14$\pm$0.18  &      26.10$\pm$0.07  &      25.21$\pm$0.07  &       ...      &       ...      & 25.08$\pm$0.33 & 24.51$\pm$0.35\\
18a  & 3:32:40.86 -27:45:46.19 & 28.76$\pm$0.09 & 26.67$\pm$0.02 & 25.83$\pm$0.01 &                &                &                &               \\
18b  & 3:32:40.86 -27:45:46.40 & 29.05$\pm$0.10 & 27.06$\pm$0.02 & 26.12$\pm$0.01 &                &                &                &               \\
20   & 3:32:45.25 -27:48:12.46 & 28.56$\pm$0.21 & 26.77$\pm$0.08 & 26.58$\pm$0.09 &       ...      &       ...      & 26.11$\pm$0.50 &  ...          \\
48ab &     ...     &          28.95$\pm$0.26  &      26.74$\pm$0.08  &      26.33$\pm$0.09  & 26.37$\pm$0.20 & 26.58$\pm$0.14 & 24.40$\pm$0.62 &  ...          \\
48a  & 3:32:45.80 -27:47:25.30 & 29.71$\pm$0.20 & 27.34$\pm$0.03 & 26.88$\pm$0.03 &                &                &                &               \\
48b  & 3:32:45.78 -27:47:25.26 & 29.69$\pm$0.23 & 27.67$\pm$0.04 & 27.34$\pm$0.05 &                &                &                &               \\
51abcde &  ...     &          27.63$\pm$0.17  &      26.21$\pm$0.08  &      26.11$\pm$0.09  &        ...     &       ...      & 25.21$\pm$0.37 &  ...          \\
51a  & 3:32:43.40 -27:46:26.87 & 28.59$\pm$0.09 & 27.35$\pm$0.04 & 27.22$\pm$0.06 &                &                &                &               \\
51b  & 3:32:43.36 -27:46:27.77 & 29.50$\pm$0.17 & 27.74$\pm$0.04 & 27.68$\pm$0.07 &                &                &                &               \\
51c  & 3:32:43.38 -27:46:27.80 & 29.11$\pm$0.11 & 27.75$\pm$0.04 & 27.61$\pm$0.06 &                &                &                &               \\
51d  & 3:32:43.31 -27:46:28.52 & 29.92$\pm$0.18 & 28.59$\pm$0.06 & 28.60$\pm$0.11 &                &                &                &               \\
51e  & 3:32:43.44 -27:46:28.52 & 31.49$\pm$0.75 & 29.46$\pm$0.14 & 29.28$\pm$0.20 &                &                &                &               \\

\enddata
\tablenotetext{1.}{Similar to Table 1, but for the IRAC-identified 
$z\approx 5$ objects. The object IDs are taken from Yan et al. (2005, in
preparation) and are independent from those of the $z\approx 6$ objects reported
in Table 1.}
\tablenotetext{2.}{This objects is detected in $K_s$ band as well: $K_s=24.62\pm0.24$ mag.}
\end{deluxetable}

\begin{deluxetable}{cccccccccc}
\tablewidth{0pc}
\tabletypesize{\scriptsize}
\tablecaption{Best-fit results for the IRAC-detected $z\approx 6$ objects\tablenotemark{1}}
\tablehead{
\colhead{ID\tablenotemark{2}} &
\colhead{Redshift\tablenotemark{3}} &
\colhead{$\tau_E$ (Gyr)\tablenotemark{4}} &
\colhead{$T_E$ (Gyr)\tablenotemark{4}} &
\colhead{$M_E$ ($M_\odot$)\tablenotemark{4}} &
\colhead{$\tau_Y$ (Myr) \tablenotemark{5}} &
\colhead{$T_Y$ (Myr)\tablenotemark{5}} &
\colhead{$M_Y$ ($M_\odot$)\tablenotemark{5}} &
\colhead{Metallicity ($Z_\odot$)} &
\colhead{$E(B-V)$}
}
\startdata

1ab  & 5.83$^*$   & 0    & 0.5 & 3.4$\times 10^{10}$ & 0 & 1.0 & 2.3$\times 10^8$ & 1 & 0 \\
5abc & $\sim$ 5.9 & 0.2  & 0.9 & 3.8$\times 10^{10}$ & 0 & 0 & 1.0$\times 10^8$ & 1 & 0 \\
7ab  & $\sim$ 5.9 & 0.01 & 0.1 & 4.7$\times 10^{9}$ & ... & ...   & ... & 1 & 0

\enddata
\tablenotetext{1.}{Parameters of the best-fit stellar population models shown in Fig. 3.}
\tablenotetext{2.}{The IDs of these multiple systems are the combination of the
IDs of their individual members.}
\tablenotetext{3.}{The asterisk indicates spectroscopic redshift, while the
leading $\sim$ sign indicates photometric redshift.}
\tablenotetext{4.}{These values are for the major, evolved component.}
\tablenotetext{5.}{These values are for the secondary, young component. If these values
are not present, it means the object can be explained by a single-component whose
parameters are given to the left.}
\end{deluxetable}

\begin{deluxetable}{cccccccccc}
\tablewidth{0pc}
\tabletypesize{\scriptsize}
\tablecaption{Best-fit results for the IRAC-detected $z\approx 5$ objects\tablenotemark{1}}
\tablehead{
\colhead{ID} &
\colhead{Redshift} &
\colhead{$\tau_E$ (Gyr)} &
\colhead{$T_E$ (Gyr)} &
\colhead{$M_E$ ($M_\odot$)} &
\colhead{$\tau_Y$ (Myr)} &
\colhead{$T_Y$ (Myr)} &
\colhead{$M_Y$ ($M_\odot$)} &
\colhead{Metallicity ($Z_\odot$)} &
\colhead{$E(B-V)$} 
}
\startdata

 6ab   & 4.65$^*$   & 0.4 & 1.3 & 2.8$\times 10^{10}$ & ... &  ...  &  ... & 1/200 & 0 \\
15ab   & 5.49$^*$   & 0.6 & 1.0 & 2.2$\times 10^{10}$ & ... &  ...  &  ... & 1 & 0 \\
48ab   & $\sim$ 5.2 &  0  & 1.0 & 1.2$\times 10^{10}$ &  0  &  10   & 2.2$\times 10^8$ & 0.4 & 0 \\
\tableline
 10    & $\sim$ 5.2 & 0.2 & 0.4 & 9.8$\times 10^{9}$  & ... & ... & ... & 1 & 0 \\
 13    & $\sim$ 4.7 & 0.4 & 1.3 & 1.8$\times 10^{10}$ & ... & ... & ... & 1/200 & 0 \\
14ab   & $\sim$ 4.5 & 0.2 & 0.7 & 2.1$\times 10^{10}$ & ... & ... & ... & 1 & 0 \\
 20    & $\sim$ 4.9 & 0.7 & 0.9 & 3.7$\times 10^{9}$  & ... & ... & ... & 1/200 & 0 \\
51abcde & $\sim$4.6 & 0.4 & 1.0 & 1.0$\times 10^{10}$ & ... & ... & ... & 1/200 & 0 \\

\enddata
\tablenotetext{1.}{Similar to Table 3, but for the $z\approx 5$ sources shown
in Fig. 5 and 7. The top part and the bottom part (separated by the horizontal
line) are for the sources with and without NICMOS measurements, respectively.}
\end{deluxetable}

\end{document}